\documentclass[reprint,superscriptaddress,amsmath,amssymb,aps,prr,longbibliography]{revtex4-1}
\usepackage[pdftex]{graphicx}
\usepackage{comment}
\usepackage{multirow}
\usepackage{url}
\usepackage[labelformat=empty,labelsep=none]{caption}
\usepackage{bm}
\usepackage{comment}
\usepackage{here}
\usepackage{color}

\begin{document}

\title{Catalog of Gamma-ray Glows during Four Winter Seasons in Japan}

\author{Y.~Wada}
\affiliation{Division of Electrical, Electronic and Infocommunications Engineering, Graduate School of Engineering, Osaka University, 2-1 Yamadaoka, Suita, Osaka 565-0871, Japan}
\affiliation{Extreme Natural Phenomena RIKEN Hakubi Research Team, Cluster for Pioneering Research, RIKEN, 2-1 Hirosawa, Wako, Saitama 351-0198, Japan}
\author{T.~Matsumoto}
\affiliation{Department of Physics, Graduate School of Science, The University of Tokyo, 7-3-1 Hongo, Bunkyo-ku, Tokyo 113-0033, Japan}
\author{T.~Enoto}
\affiliation{Extreme Natural Phenomena RIKEN Hakubi Research Team, Cluster for Pioneering Research, RIKEN, 2-1 Hirosawa, Wako, Saitama 351-0198, Japan}
\author{K.~Nakazawa}
\affiliation{Kobayashi-Maskawa Institute for the Origin of Particles and the Universe, Nagoya University, Furo-cho, Chikusa-ku, Nagoya, Aichi 464-8601, Japan}
\author{T.~Yuasa}
\affiliation{Block 4B, Boon Tiong Road, Singapore 165004, Singapore}
\author{Y.~Furuta}
\affiliation{CLEAR-PULSE Co., LTD., 6-25-27 Chuo, Ota-ku, Tokyo 143-0024, Japan}
\author{D.~Yonetoku}
\affiliation{School of Mathematics and Physics, College of Science and Engineering, Kanazawa University, Kakuma-cho, Kanazawa, Ishikawa 920-1192, Japan}
\author{T.~Sawano}
\affiliation{Advanced Research Center for Space Science and Technology, Institute of Science and Engineering, Kanazawa University, Kakuma-cho, Kanazawa, Ishikawa 920-1192, Japan}
\author{G.~Okada}
\affiliation{Co-creative Research Center of Industrial Science and Technology (CIST), Kanazawa Institute of Technology, 3-1 Yatsukaho, Hakusan, Ishikawa 924-0838, Japan}
\author{H.~Nanto}
\affiliation{Co-creative Research Center of Industrial Science and Technology (CIST), Kanazawa Institute of Technology, 3-1 Yatsukaho, Hakusan, Ishikawa 924-0838, Japan}
\author{S.~Hisadomi}
\affiliation{Division of Particle and Astrophysical Science, Graduate School of Science, Nagoya University, Furo-cho, Chikusa-ku, Nagoya, Aichi 464-8601, Japan}
\author{Y.~Tsuji}
\affiliation{Division of Particle and Astrophysical Science, Graduate School of Science, Nagoya University, Furo-cho, Chikusa-ku, Nagoya, Aichi 464-8601, Japan}
\author{G.~S.~Diniz}
\affiliation{Extreme Natural Phenomena RIKEN Hakubi Research Team, Cluster for Pioneering Research, RIKEN, 2-1 Hirosawa, Wako, Saitama 351-0198, Japan}
\author{K.~Makishima}
\affiliation{Department of Physics, Graduate School of Science, The University of Tokyo, 7-3-1 Hongo, Bunkyo-ku, Tokyo 113-0033, Japan}
\affiliation{High Energy Astrophysics Laboratory, Nishina Center for Accelerator-Based Science, RIKEN, 2-1 Hirosawa, Wako, Saitama 351-0198, Japan}
\affiliation{Kavli Institute for the Physics and Mathematics of the Universe, The University of Tokyo, 5-1-5 Kashiwa-no-ha, Kashiwa, Chiba 277-8683, Japan}
\author{H.~Tsuchiya}
\affiliation{Nuclear Science and Engineering Center, Japan Atomic Energy Agency, 2-4 Shirane Shirakata, Tokai-mura, Naka-gun, Ibaraki 319-1195, Japan}

\begin{abstract}

	In 2015 the Gamma-Ray Observation of Winter Thunderstorms (GROWTH) collaboration launched a mapping observation campaign 
	for high-energy atmospheric phenomena related to thunderstorms and lightning discharges.
	This campaign has developed a detection network of gamma rays with up to 10 radiation monitors installed in Kanazawa and Komatsu cities, Ishikawa Prefecture, Japan,
	where low-charge-center winter thunderstorms frequently occur. During four winter seasons from October 2016 to April 2020, in total 70 gamma-ray glows, 
	minute-lasting bursts of gamma rays originating from thunderclouds, were detected. Their average duration is 58.9~sec.  Among the detected events, 77\% were observed in nighttime. 
	The gamma-ray glows can be classified into temporally-symmetric, temporally-asymmetric, and lightning-terminated types based on their count-rate histories.
	An averaged energy spectrum of the gamma-ray glows is well fitted with a power-law function with an exponential cutoff, whose photon index, cutoff energy, and flux are 
	$0.613\pm0.009$, $4.68\pm0.04$~MeV, and $(1.013\pm0.003)\times10^{-5}$~erg~cm$^{-2}$~s$^{-1}$ (0.2-20.0~MeV), respectively.
	The present paper provides the first catalog of gamma-ray glows and their statistical analysis detected during winter thunderstorms in the Kanazawa and Komatsu areas.

\end{abstract}

\maketitle


\begin{figure*}[th]
	\begin{center}
	\includegraphics[width=0.8\hsize]{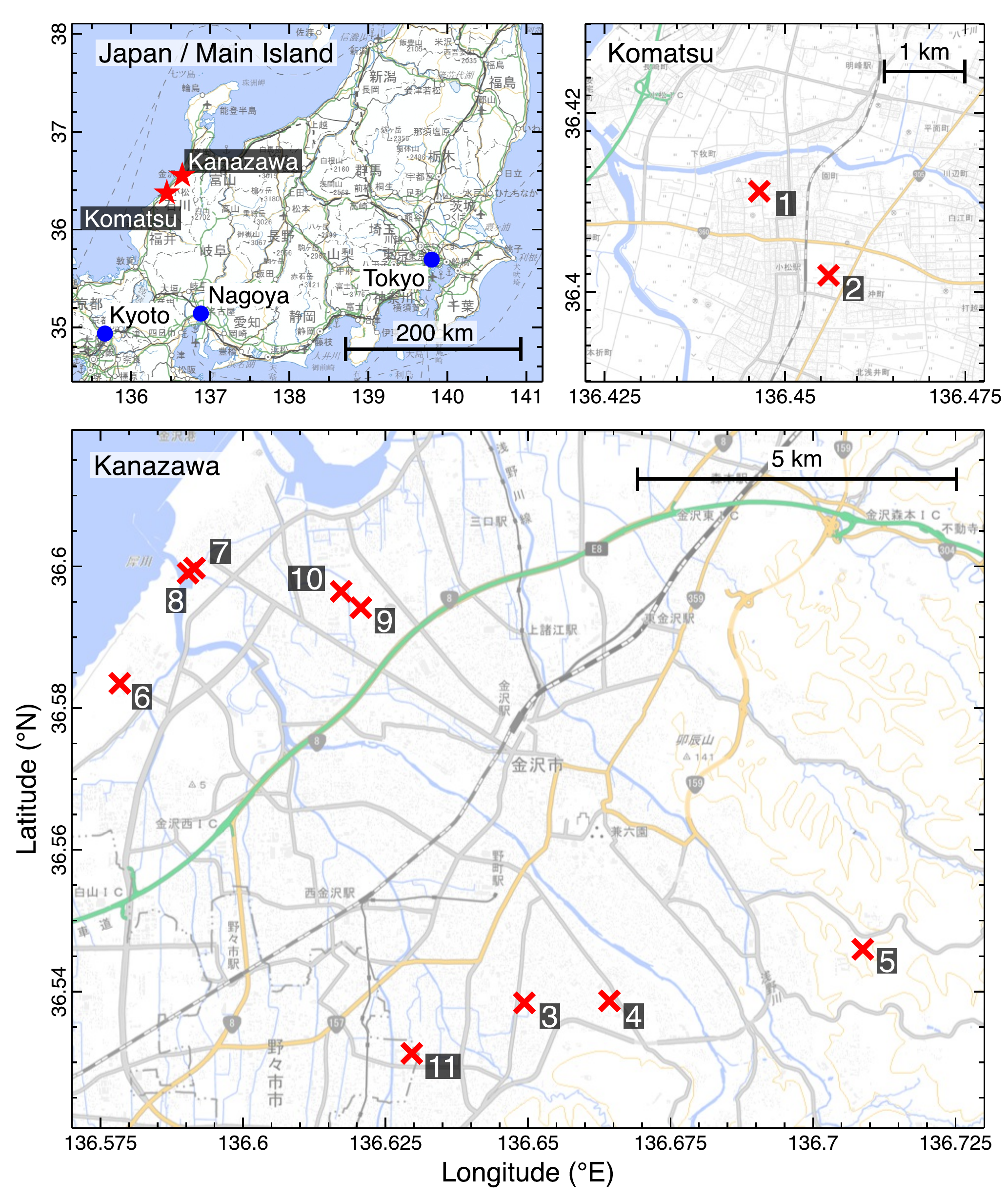}
	\caption{Figure~\ref{fig:map}: Maps of observation sites. The red-cross markers indicate the position of radiation monitors with their ID numbers.
	The background maps are provided by Geospatial Information Authority of Japan.}
	\label{fig:map}
	\end{center}
\end{figure*}

\begin{figure*}[tbh]
	\begin{center}
	\includegraphics[width=0.8\hsize]{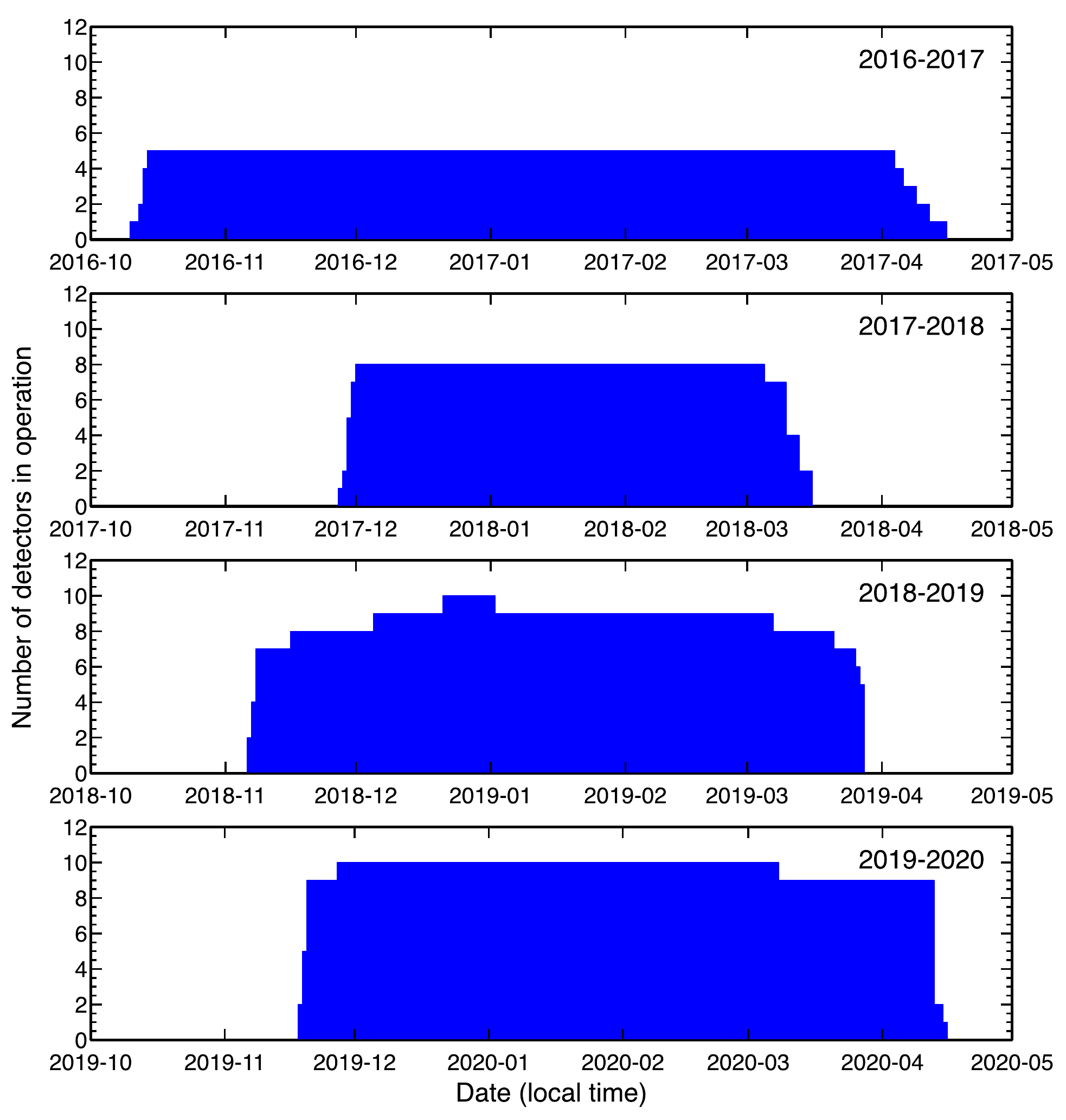}
	\caption{Figure~\ref{fig:operation}: Operation histories of the number of radiation detectors during the four winter seasons.}
	\label{fig:operation}
	\end{center}
\end{figure*}

\setlength{\tabcolsep}{2.5mm}
\begin{table*}[tbh]
\caption{Table~\ref{tab:detector}: Specifications and observation periods of radiation monitors.}
\begin{tabular}{c c c c c c c c}
\hline
\multirow{2}{*}{ID} & \multirow{2}{*}{Site} & \multirow{2}{*}{Crystal} & \multicolumn{4}{c}{Operation period} \\
& & & FY2016 & FY2017 & FY2018 & FY2019 \\ \hline
\multirow{2}{*}{1} & Komatsu High School & \multirow{2}{*}{BGO} & 2016.10.13-- & 2017.11.28-- & 2018.11.08-- & 2019.11.19-- \\
& (36.411$^{\circ}$N, 136.446$^{\circ}$E) & & 2017.04.03 & 2018.03.15 & 2019.03.27 & 2020.04.12 \\ \hline
\multirow{2}{*}{2} & Science Hills Komatsu & \multirow{2}{*}{BGO} & 2016.10.14-- & 2017.11.30-- & 2018.11.08-- & 2019.11.19-- \\
& (36.402$^{\circ}$N, 136.456$^{\circ}$E) & & 2017.04.05 & 2018.03.15 & 2019.03.27 & 2020.04.12 \\ \hline
\multirow{2}{*}{3} & Kanazawa Izumigaoka High School & \multirow{2}{*}{BGO} & 2016.10.13-- & 2017.11.27-- & 2018.11.07-- & 2019.11.18-- \\
& (36.538$^{\circ}$N, 136.649$^{\circ}$E) & & 2017.04.08 & 2018.03.04 & 2019.03.20 & 2020.04.12 \\ \hline
\multirow{2}{*}{4} & Kanazawa University High School & \multirow{2}{*}{BGO} & 2016.10.12-- & 2017.11.30-- & 2018.11.16-- & 2019.11.18-- \\
& (36.539$^{\circ}$N, 136.664$^{\circ}$E) & & 2017.04.11 & 2018.03.12 & 2019.03.25 & 2020.04.12 \\ \hline
\multirow{2}{*}{5} & Kanazawa University & \multirow{2}{*}{BGO} & 2016.10.10-- & 2017.12.01-- & 2018.11.08-- & 2019.11.20-- \\
& (36.546$^{\circ}$N, 136.709$^{\circ}$E) & & 2017.04.15 & 2018.03.12 & 2019.03.26 & 2020.04.12 \\ \hline
\multirow{2}{*}{6} & Ishikawa Plating Industry Senkoji Plant & \multirow{2}{*}{CsI} &  & 2017.11.29-- & 2018.11.06-- & 2019.11.20-- \\
& (36.583$^{\circ}$N, 136.578$^{\circ}$E) & &  & 2018.03.09 & 2019.03.27 & 2020.04.12 \\ \hline
\multirow{2}{*}{7} & Ishikawa Plating Industry Head Quarter & \multirow{2}{*}{CsI} &  & 2017.11.29-- &  & 2019.11.20-- \\
& (36.6$^{\circ}$N, 136.591$^{\circ}$E) & &  & 2018.03.09 &  & 2020.04.12 \\ \hline
\multirow{2}{*}{8} & Ishikawa Plating Industry Second Plant & \multirow{2}{*}{CsI} &  & 2017.11.29-- & 2018.11.06-- &  \\
& (36.599$^{\circ}$N, 136.59$^{\circ}$E) & &  & 2018.03.09 & 2019.03.27 &  \\ \hline
\multirow{2}{*}{9} & Industrial Research Institute of Ishikawa & \multirow{2}{*}{CsI} &  &  & 2018.12.05-- & 2019.11.19-- \\
& (36.594$^{\circ}$N, 136.621$^{\circ}$E) & &  &  & 2019.03.06 & 2020.04.15 \\ \hline
\multirow{2}{*}{10} & Kanazawa Nishi High School & \multirow{2}{*}{CsI} &  &  & 2018.11.07-- & 2019.11.20-- \\
& (36.596$^{\circ}$N, 136.617$^{\circ}$E) & &  &  & 2019.01.01 & 2020.03.07 \\ \hline
\multirow{2}{*}{11} & Kanazawa Institute of Technology & \multirow{2}{*}{CsI} &  &  & 2018.12.21-- & 2019.11.27-- \\
& (36.531$^{\circ}$N, 136.63$^{\circ}$E) & &  &  & 2019.03.27 & 2020.04.14 \\ \hline
\end{tabular}
\label{tab:detector}
\end{table*}
\setlength{\tabcolsep}{1mm}
	
\section{Introduction}
	Strong electric fields inside thunderclouds have been recently recognized as a particle accelerator in nature.
	Gamma-ray glows, also called thunderstorm ground enhancements (TGEs) when detected at the ground, 
	are minute-lasting high-energy phenomena in the atmosphere caused by strong electric fields \cite{McCarthy_1985,Torii_2002,Tsuchiya_2007,Chilingarian_2010}.
	Electrons accelerated to a relativistic energy in an electric field emit bremsstrahlung photons by colliding with atmospheric atoms.
	The energy spectrum of gamma-ray glows typically extend up to tens of MeV \cite{Tsuchiya_2011,Wada_2018}, 
	and their observed duration ranges from seconds to tens of minutes \cite{Kelley_2015,Tsuchiya_2011,Tsuchiya_2012,Chilingarian_2019b}.

	Since the first detection by an F-106 aircraft \cite{Parks_1981,McCarthy_1985}, 
	gamma-ray glows have been detected by experiments onboard balloon \cite{Moore_2001} and aircraft \cite{Kelley_2015,Ostgaard_2019}, 
	or those located on mountain  \cite{Chilingarian_2010,Torii_2009} and at sea-level \cite{Torii_2002,Tsuchiya_2007,Kuroda_2016}.
	Thank to the recent advances in detectors, aircraft experiments have performed precise measurements of gamma-ray glows
	inside or in the vicinity of thunderclouds \cite{Kelley_2015,Kochkin_2017,Ostgaard_2019,Kochkin_2021}.
	At the same time, ground-based experiments have also greatly contributed to gamma-ray glow studies.
	
	Mountain experiments are ones of the important ways to study gamma-ray glows/TGEs at the ground.
	Electric fields of thunderclouds are usually formed at an altitude of 3~km or higher in summer. 
	Gamma rays coming out from the electric-field regions cannot reach the sea level as they are absorbed by the atmosphere.
	Therefore, mountain experiments at an altitude of 2--3~km or higher are necessary for glow studies during summertime thunderstorms.
	So far cosmic-ray observatories have observed gamma-ray glows; the Aragats Space Environmental Center in Armenia \cite{Chilingarian_2010,Chilingarian_2012,Chilingarian_2017}, 
	Yangbajing in Tibet, China \cite{Tsuchiya_2012}, Mt. Norikura in Japan \cite{Tsuchiya_2009}, Tien Shan in Kazakhstan \cite{Shepetov_2021}, 
	as well as the weather station at Mt. Fuji in Japan \cite{Torii_2009}.
	In particular the Aragats Space Environmental Center have observed the largest number of TGEs in the world \cite{Chilingarian_2019b}.

\begin{table*}[p]
\caption{Table~\ref{tab:event1}: Event list of gamma-ray glows}
\begin{tabular}{c c c c c c c c c c c c c}
\hline
& & & & & Num. of & & Energy & Timing & \multicolumn{2}{c}{Wind$^{c}$} & & \\
No. & Detection time & Site & Type$^{a}$ & Signif.$^{b}$ & photons & Duration & range & accuracy & speed & direction & Temp. & Pub.$^{d}$ \\
& (JST) & & & & & (sec) & (MeV) & & (m~s$^{-1}$) & (degree) & ($^{\circ}$C) & \\ \hline
1 & 2016/12/08 00:13:39 & 5 & S & 10.6 & 491$\pm$28 & 61 & 0.3--15 & $\pm$0.5~s & 9.5$\pm$1.2 & 260 & 7.2 & \cite{Wada_2021} \\
2 & 2016/12/08 02:56:24 & 1 & S & 11.5 & 807$\pm$38 & 118 & 0.7--16 & $\pm$0.5~s & 10.9$\pm$1.2 & 300 & 5.9 & \cite{Yuasa_2020} \\
3 & 2016/12/08 02:58:21 & 2 & S & 43.8 & 3421$\pm$38 & 98 & 0.4--14 & $\pm$0.5~s & 10.9$\pm$1.2 & 300 & 5.9 & \cite{Yuasa_2020} \\
4 & 2016/12/09 16:29:03 & 3 & T & 9.2 & 326$\pm$24 & 43.5 & 0.7--14 & $\pm$0.5~s & 18.1$\pm$1.4 & 260 & 9.7 &  \\
5 & 2017/01/13 01:43:38 & 5 & S & 10.0 & 335$\pm$28 & 46 & 0.3--15 & $\pm$0.5~s & 20.1$\pm$2.2 & 260 & 5.7 & \cite{Wada_2021} \\
6 & 2017/01/13 05:05:22 & 5 & A & 184.2 & 5236$\pm$28 & 42 & 0.3--14 & $\pm$0.5~s & 16.1$\pm$1.5 & 270 & 6.1 & \cite{Wada_2021} \\
7 & 2017/01/15 05:30:47 & 5 & S & 12.0 & 522$\pm$28 & 56 & 0.3--14 & $\pm$0.5~s & 16.7$\pm$1 & 310 & -0.4 & \cite{Wada_2021} \\
8 & 2017/02/06 05:06:46 & 5 & S & 5.4 & 228$\pm$33 & 86 & 0.3--15 & $\pm$0.5~s & 16.5$\pm$1.1 & 260 & 6.8 & \cite{Wada_2021} \\
9 & 2017/02/06 05:10:33 & 5 & S & 5.3 & 67$\pm$38 & 101 & 0.3--15 & $\pm$0.5~s & 16.5$\pm$1.1 & 260 & 6.8 & \cite{Wada_2021} \\
10 & 2017/12/05 03:27:07 & 1 & S & 9.7 & 295$\pm$25 & 45 & 0.4--25 & $\pm$1~s & 13.2$\pm$1.3 & 250 & 6.8 &  \\
11 & 2017/12/05 03:27:39 & 2 & S & 8.9 & 168$\pm$26 & 35 & 0.4--23 & $\pm$1~$\mu$s & 13.2$\pm$1.3 & 250 & 6.8 &  \\
12 & 2017/12/05 11:51:43 & 6 & S & 15.5 & 454$\pm$24 & 38 & 0.4--24 & $\pm$1~$\mu$s & 20.2$\pm$1.4 & 240 & 5 &  \\
13 & 2017/12/05 12:07:13 & 1 & S & 74.1 & 2327$\pm$28 & 44 & 0.4--24 & $\pm$1~s & 21.6$\pm$1.2 & 240 & 4 &  \\
14 & 2017/12/05 18:33:38 & 5 & A & 14.0 & 1030$\pm$37 & 108 & 0.4--23 & $\pm$1~$\mu$s & 14.8$\pm$1 & 260 & 6.6 & \cite{Wada_2020,Wada_2021} \\
15 & 2017/12/11 17:49:01 & 3 & T & 24.4 & 936$\pm$23 & 45 & 0.4--20 & unknown & 21.6$\pm$1.1 & 250 & 4.9 &  \\
16 & 2017/12/11 17:49:01 & 4 & T & 98.9 & 1388$\pm$19 & 25.5 & 0.4--21 & $\pm$1~$\mu$s & 21.6$\pm$1.1 & 250 & 4.9 &  \\
17 & 2017/12/11 17:52:05 & 5 & A & 16.0 & 632$\pm$38 & 121 & 0.4--23 & $\pm$1~$\mu$s & 21.6$\pm$1.1 & 250 & 4.9 & \cite{Wada_2021} \\
18 & 2017/12/11 18:17:39 & 5 & S & 4.7 & N/A & N/A & 0.4--23 & $\pm$1~$\mu$s & 21$\pm$1.2 & 250 & 4 &  \\
19 & 2017/12/26 08:10:12 & 3 & S & 124.3 & 3305$\pm$28 & 34.5 & 0.4--20 & unknown & 20.6$\pm$1.4 & 260 & 5.8 &  \\
20 & 2017/12/26 08:11:10 & 4 & S & 91.8 & 3239$\pm$27 & 39 & 0.4--21 & $\pm$1~$\mu$s & 20.6$\pm$1.4 & 260 & 5.8 &  \\
21 & 2018/01/03 03:19:36 & 3 & S & 11.8 & 251$\pm$21 & 24.5 & 0.4--20 & unknown & 11.8$\pm$1.2 & 290 & 2.7 &  \\
22 & 2018/01/09 23:41:36 & 2 & A & 112.7 & 6496$\pm$37 & 92 & 0.4--22 & $\pm$1~$\mu$s & 15.9$\pm$1.4 & 250 & 1.8 &  \\
23 & 2018/01/10 02:36:55 & 1 & S & 19.5 & 468$\pm$23 & 33 & 0.4--24 & $\pm$1~s & 20.4$\pm$1.3 & 250 & 1.6 &  \\
24 & 2018/01/10 02:53:54 & 3 & S & 123.7 & 3458$\pm$25 & 35.5 & 0.4--20 & $<$1~ms & 19.3$\pm$1.4 & 250 & 2 & \cite{Wada_2019_commphys,Wada_2020} \\
25 & 2018/01/10 02:54:50 & 4 & TF & 770.4 & 9012$\pm$19 & 18 & 0.4--20 & $\pm$1~$\mu$s & 19.3$\pm$1.4 & 250 & 2 & \cite{Wada_2019_commphys,Wada_2020} \\
26 & 2018/01/10 02:57:56 & 3 & S & 699.1 & 24081$\pm$34 & 47 & 0.4--20 & $\pm$1~s & 19.3$\pm$1.4 & 250 & 2.2 &  \\
27 & 2018/01/10 07:34:56 & 1 & A & 9.3 & 419$\pm$31 & 103 & 0.4--24 & $\pm$1~s & 16.1$\pm$1.2 & 250 & 1.6 &  \\
28 & 2018/01/10 11:59:10 & 6 & A & 3.3 & N/A & N/A & 0.4--25 & $\pm$1~$\mu$s & 13.8$\pm$1.6 & 250 & 2.2 &  \\
29 & 2018/01/10 12:02:25 & 6 & A & 27.4 & 1168$\pm$21 & 48 & 0.4--25 & $\pm$1~$\mu$s & 13.8$\pm$1.6 & 250 & 2.2 &  \\
30 & 2018/01/11 10:40:13 & 6 & A & 34.0 & 1075$\pm$29 & 53 & 0.4--25 & $\pm$1~$\mu$s & 18.2$\pm$1.3 & 240 & -0.4 &  \\
31 & 2018/01/13 12:01:01 & 8 & S & 13.3 & 368$\pm$21 & 40.5 & 0.4--26 & $\pm$1~$\mu$s & 10.8$\pm$1.5 & 260 & 1.2 &  \\
32 & 2018/01/13 12:01:04 & 7 & S & 15.2 & 511$\pm$24 & 53 & 0.4--27 & $\pm$1~$\mu$s & 10.8$\pm$1.5 & 260 & 1.2 &  \\
33 & 2018/01/23 13:29:06 & 2 & S & 5.7 & 265$\pm$36 & 84 & 0.4--23 & $\pm$1~$\mu$s & 25.6$\pm$1.3 & 250 & 3.2 &  \\
34 & 2018/02/04 03:50:31 & 6 & S & 8.2 & 226$\pm$36 & 97 & 0.4--24 & $\pm$1~$\mu$s & 13.8$\pm$1.5 & 250 & 1.1 &  \\
35 & 2018/12/08 20:53:18 & 5 & S & 5.1 & 152$\pm$36 & 46 & 0.4--23 & $\pm$1~$\mu$s & 14$\pm$1.3 & 250 & 5.1 &  \\
36 & 2018/12/08 21:09:21 & 3 & S & 13.2 & 776$\pm$33 & 93.5 & 0.4--20 & $\pm$1~s & 14$\pm$1.3 & 250 & 5.5 &  \\
37 & 2018/12/08 21:10:57 & 4 & S & 6.7 & 503$\pm$42 & 161.5 & 0.4--21 & $\pm$1~$\mu$s & 14$\pm$1.3 & 250 & 5.5 &  \\
38 & 2018/12/17 18:51:30 & 10 & T & 3.9 & N/A & N/A & 0.4--22 & $\pm$1~$\mu$s & 10.9$\pm$1.3 & 270 & 7.7 &  \\
39 & 2018/12/17 18:54:11 & 9 & T & 9.1 & 95$\pm$11 & 12 & 0.4--18 & $\pm$1~$\mu$s & 10.9$\pm$1.3 & 270 & 7.7 &  \\
40 & 2018/12/17 18:54:11 & 10 & T & 13.6 & 273$\pm$15 & 23.5 & 0.4--22 & $\pm$1~$\mu$s & 10.9$\pm$1.3 & 270 & 7.7 &  \\
\hline
\multicolumn{13}{l}{$^{a}$ S, A, and T mean a gamma-ray glow with temporarily-symmetric count-rate variation, }\\
\multicolumn{13}{l}{\;\;\;with temporarily-asymmetric count-rate variation, and a glow terminated with a lightning discharge, }\\
\multicolumn{13}{l}{\;\;\;respectively. TF means a glow terminated with a downward TGF besides a lightning discharge.} \\
\multicolumn{13}{l}{$^{b}$ The significance of each event calculated for the event scan.} \\
\multicolumn{13}{l}{$^{c}$ The wind direction is where the wind comes from in clockwise. See the main text.} \\
\multicolumn{13}{l}{$^{d}$ Publication references if it is a previously-published event.} \\
\end{tabular}
\label{tab:event1}
\end{table*}

\begin{table*}[tbh]
\caption{(continued)}
\begin{tabular}{c c c c c c c c c c c c c}
\hline
& & & & & Num. of & & Energy & Timing & \multicolumn{2}{c}{Wind} & & \\
No. & Detection time & Site & Type & Signif. & photons & Duration & range & accuracy & speed & direction & Temp. & Pub. \\
& (JST) & & & & & (sec) & (MeV) & & (m~s$^{-1}$) & (degree) & ($^{\circ}$C) & \\ \hline
41 & 2018/12/17 19:15:03 & 10 & S & 7.4 & 383$\pm$28 & 69 & 0.4--22 & $\pm$1~$\mu$s & 10.6$\pm$1.2 & 270 & 5.9 &  \\
42 & 2018/12/17 19:15:16 & 9 & S & 18.2 & 678$\pm$24 & 51.5 & 0.4--18 & $\pm$1~$\mu$s & 10.6$\pm$1.2 & 270 & 5.9 &  \\
43 & 2018/12/17 23:15:34 & 3 & S & 20.6 & 1024$\pm$29 & 63 & 0.4--20 & $\pm$1~s & 11.3$\pm$1.3 & 280 & 8.4 &  \\
44 & 2018/12/17 23:17:04 & 4 & S & 24.3 & 1399$\pm$33 & 89.5 & 0.4--22 & $\pm$1~$\mu$s & 11.3$\pm$1.3 & 280 & 8.4 &  \\
45 & 2018/12/18 23:40:06 & 8 & T & 6.8 & 86$\pm$13 & 15.5 & 0.4--24 & $\pm$1~$\mu$s & 13.5$\pm$1.3 & 270 & 8.1 &  \\
46 & 2018/12/18 23:42:05 & 8 & S & 53.6 & 2117$\pm$27 & 56.5 & 0.4--24 & $\pm$1~$\mu$s & 13.5$\pm$1.3 & 270 & 8.1 &  \\
47 & 2018/12/18 23:42:56 & 9 & T & 38.2 & 515$\pm$16 & 25.5 & 0.4--18 & $\pm$1~$\mu$s & 13.5$\pm$1.3 & 270 & 8.1 &  \\
48 & 2018/12/18 23:42:56 & 10 & T & 9.8 & 131$\pm$17 & 26.5 & 0.4--22 & $\pm$1~$\mu$s & 13.5$\pm$1.3 & 270 & 8.1 &  \\
49 & 2018/12/18 23:49:40 & 5 & T & 10.5 & 147$\pm$14 & 17 & 0.4--24 & $\pm$1~$\mu$s & 11.9$\pm$1 & 270 & 6.6 & \cite{Wada_2021} \\
50 & 2018/12/18 23:51:48 & 5 & T & 311.0 & 7351$\pm$23 & 32 & 0.4--24 & $\pm$1~$\mu$s & 11.9$\pm$1 & 270 & 6.6 & \cite{Wada_2021} \\
51 & 2019/01/22 00:51:33 & 8 & T & 11.9 & 331$\pm$19 & 31 & 0.4--24 & $\pm$1~$\mu$s & 14.2$\pm$1.2 & 280 & 5.6 &  \\
52 & 2019/01/22 01:03:48 & 5 & A & 38.5 & 1678$\pm$35 & 69 & 0.4--23 & $\pm$1~$\mu$s & 13.6$\pm$1.1 & 270 & 4.2 & \cite{Wada_2021} \\
53 & 2019/01/25 20:43:43 & 6 & S & 10.6 & 293$\pm$25 & 39.5 & 0.4--25 & $\pm$1~$\mu$s & 16.9$\pm$1.2 & 260 & 2.8 &  \\
54 & 2019/01/25 20:44:30 & 8 & S & 8.2 & 324$\pm$28 & 51 & 0.4--24 & $\pm$1~$\mu$s & 16.9$\pm$1.2 & 260 & 2.8 &  \\
55 & 2020/01/13 02:03:23 & 3 & T & 70.0 & 1034$\pm$18 & 19.5 & 0.4--20 & $\pm$1~$\mu$s & 16.4$\pm$1.4 & 240 & 6.2 & \cite{Hisadomi_2021} \\
56 & 2020/01/13 02:03:23 & 4 & T & 32.7 & 354$\pm$20 & 14 & 0.5--22 & $<$1~ms & 16.4$\pm$1.4 & 240 & 6.2 & \cite{Hisadomi_2021} \\
57 & 2020/01/13 02:05:35 & 3 & A & 131.4 & 3339$\pm$29 & 34 & 0.4--20 & $\pm$1~$\mu$s & 16.4$\pm$1.4 & 240 & 6.2 & \cite{Hisadomi_2021} \\
58 & 2020/01/13 02:06:59 & 4 & TF & 111.9 & 5316$\pm$32 & 59 & 0.5--22 & $<$1~ms & 16.4$\pm$1.4 & 240 & 6.2 & \cite{Hisadomi_2021} \\
59 & 2020/02/17 20:52:56 & 11 & TF & 7.8 & 112$\pm$17 & 27 & 0.7--15 & $\pm$1~$\mu$s & 15.7$\pm$1.2 & 260 & 2.5 &  \\
60 & 2020/02/17 21:00:14 & 5 & A & 61.6 & 3605$\pm$44 & 111 & 0.4--23 & $\pm$1~$\mu$s & 15.7$\pm$1.2 & 260 & 2.4 &  \\
61 & 2020/02/17 21:47:11 & 11 & A & 3.9 & N/A & N/A & 0.7--15 & $\pm$1~$\mu$s & 15$\pm$1.2 & 250 & 2.6 &  \\
62 & 2020/02/17 21:49:40 & 3 & S & 10.3 & 501$\pm$26 & 63 & 0.4--20 & $\pm$1~$\mu$s & 15$\pm$1.2 & 250 & 2.6 &  \\
63 & 2020/02/17 21:49:46 & 11 & A & 9.4 & 634$\pm$32 & 86 & 0.7--15 & $\pm$1~$\mu$s & 15$\pm$1.2 & 250 & 2.6 &  \\
64 & 2020/02/17 21:51:17 & 4 & S & 10.9 & 805$\pm$29 & 73 & 0.5--20 & $\pm$1~$\mu$s & 15$\pm$1.2 & 250 & 2.6 &  \\
65 & 2020/02/17 21:52:17 & 3 & A & 84.2 & 3958$\pm$35 & 86 & 0.4--20 & $\pm$1~$\mu$s & 15$\pm$1.2 & 250 & 2.6 &  \\
66 & 2020/02/17 21:53:54 & 4 & A & 99.1 & 4023$\pm$32 & 49.5 & 0.5--20 & $\pm$1~$\mu$s & 15$\pm$1.2 & 250 & 2.6 &  \\
67 & 2020/02/18 01:29:17 & 10 & A & 86.3 & 4771$\pm$33 & 89.5 & 0.3--15 & $\pm$1~$\mu$s & 12.7$\pm$1.3 & 260 & 1.1 &  \\
68 & 2020/02/18 01:30:42 & 9 & T & 60.2 & 3491$\pm$31 & 110.5 & 0.4--17 & $\pm$1~$\mu$s & 12.7$\pm$1.3 & 260 & 1.1 &  \\
69 & 2020/02/18 01:35:23 & 2 & A & 87.0 & 5355$\pm$41 & 120 & 0.4--22 & $\pm$1~$\mu$s & 13.6$\pm$1.3 & 260 & 1.9 &  \\
70 & 2020/03/16 06:28:10 & 5 & T & 41.2 & 1722$\pm$26 & 58 & 0.4--23 & $\pm$1~$\mu$s & 13$\pm$1.2 & 280 & 4.1 &  \\
\hline
\end{tabular}
\label{tab:event2}
\end{table*}

\begin{figure*}[tb]
	\begin{center}
	\includegraphics[width=0.8\hsize]{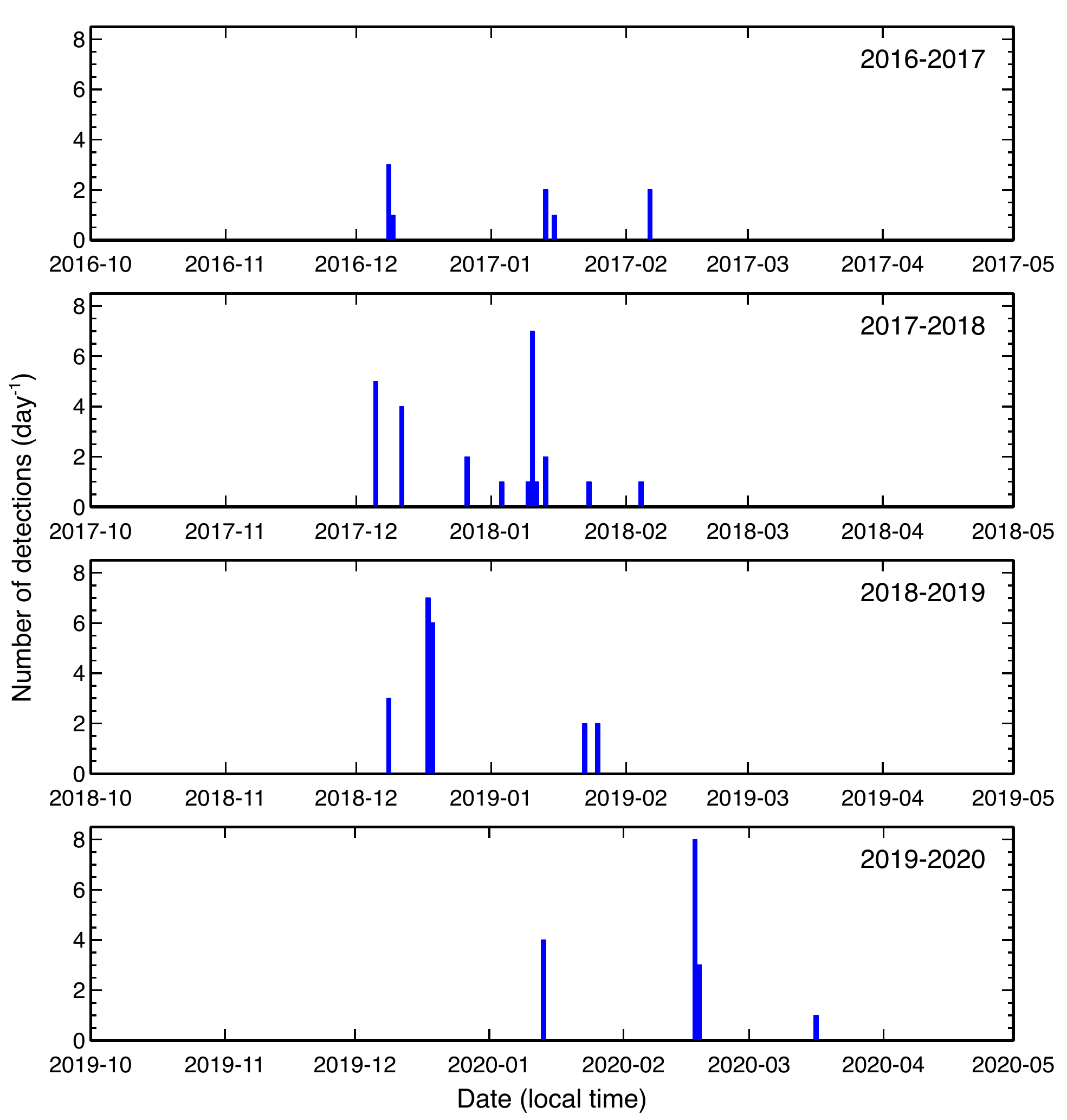}
	\caption{Figure~\ref{fig:detection}: Time histories of the number of gamma-ray glow detections per a day.}
	\label{fig:detection}
	\end{center}
\end{figure*}

\begin{figure*}[ttb]
	\begin{center}
	\includegraphics[width=0.8\hsize]{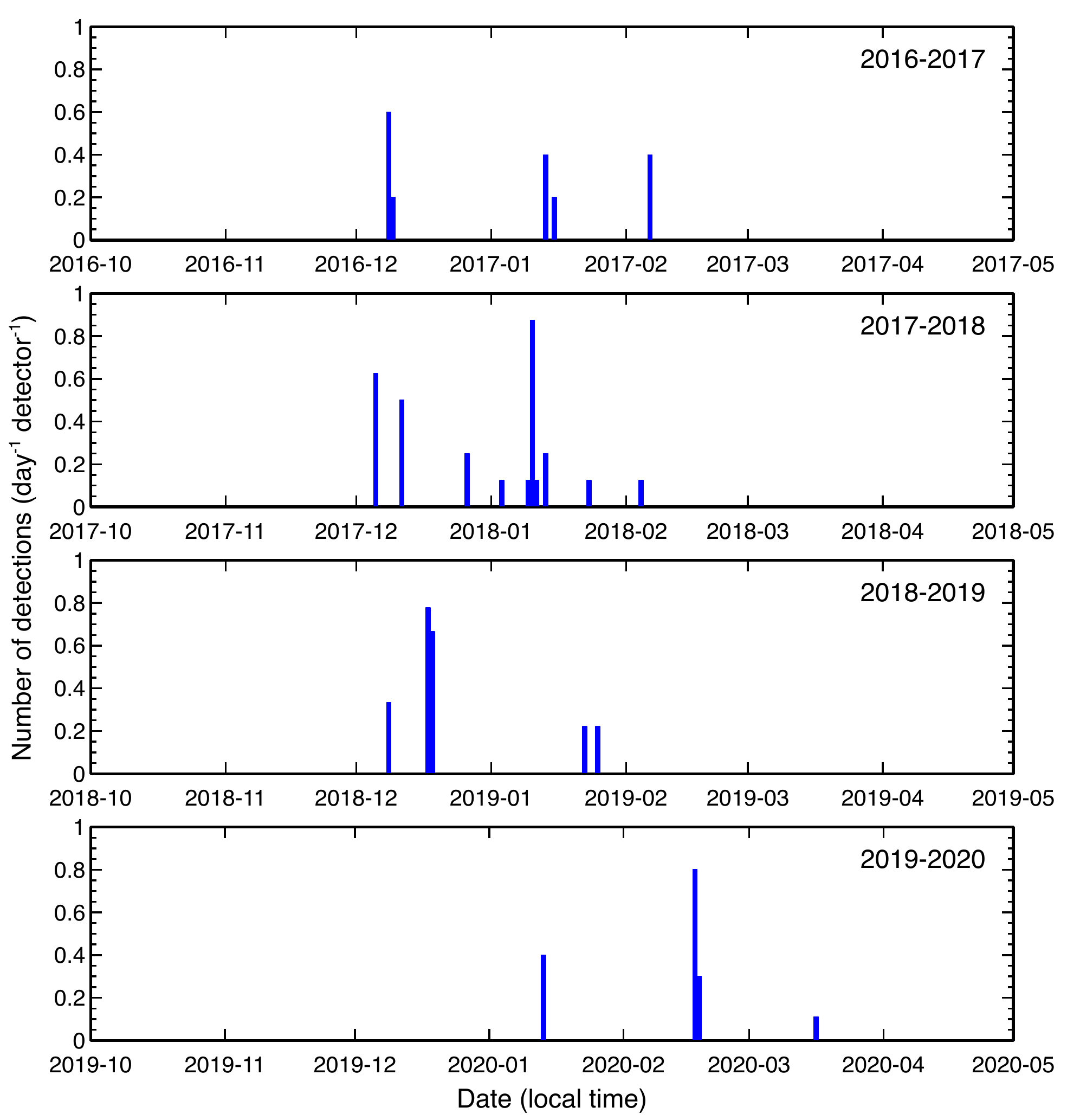}
	\caption{Figure~\ref{fig:ratio}: Time histories of the number of gamma-ray glow detections per a day, normalized by the number of radiation detectors in operaion.}
	\label{fig:ratio}
	\end{center}
\end{figure*}	
	
	The other opportunities to detect gamma-ray glows at ground level are winter thunderstorms in Japan.
	When dried and cold seasonal wind comes from a Siberian cold air mass over the Tsushima warm current, 
	an active ascending air current, then a thunderstorm can be formed in the north-west coast of the Japan main island called the "Hokuriku" region \cite{Kitagawa_1994,Rakov_2003}.
	One of the important characteristics of these winter thunderstorms distinguished from summer ones is low charge center.
	Precipitation particles that contribute to charging (i.e. ice crystals, graupels) are produced at an altitude whose temperature ranges from $-10^{\circ}$C to $0^{\circ}$C or higher \cite{Takahashi_1978,Williams_1989}.
	It typically corresponds to $\sim$3~km for summer thunderstorms, but less than 1~km for winter ones \cite{Goto_Narita_1992}.
	Therefore, gamma rays emitted from winter thunderclouds can be detected even at sea-level.
	High-energy emission form winter thunderstorms in Japan has been discovered unexpectedly and investigated by radiations monitors in nuclear power stations \cite{Torii_2002,Torii_2011}.
	Following these findings, we have performed a ground-based radiation measurement called the Gamma-Ray Observation of Winter Thunderclouds (GROWTH) experiment 
	in the Hokuriku region since 2006 \cite{Tsuchiya_2007,Tsuchiya_2013,Umemoto_2016,Enoto_2017,Wada_2018,Yuasa_2020}.
	
	Gamma-ray glows are thought to originate from strong electric fields formed within thunderclouds, not directly from lightning discharges.
	On the other hand, they are often quenched by nearby discharges since lightning currents could discharge the corresponding electric fields \cite{McCarthy_1985,Tsuchiya_2013,Kelley_2015,Chilingarian_2017,Wada_2018}. 
	More rare cases of the association between gamma-ray glows and lightning discharges are terrestrial gamma-ray flashes (TGFs).
	TGFs are instantaneous gamma-ray emission coincident with lightning discharges \cite{Fishman_1994,Smith_2005,Tavani_2011,Ostgaard_2019b}. 
	Their duration is typically a few hundreds of microseconds \cite{Foley_2014,Ostgaard_2019b}, 
	and their energy spectrum also extends up to $>$20~MeV \cite{Smith_2005,Marisaldi_2010,Mailyan_2016}. 
	TGFs have been detected mainly by space-borne detectors such as CGRO \cite{Fishman_1994}, 
	RHESSI \cite{Smith_2005}, AGILE \cite{Tavani_2011}, Fermi \cite{Briggs_2010}, and ASIM \cite{Neubert_2019b}.
	While the TGFs observed from space are upward-going, downward TGFs have been recently detected by ground-based experiments \cite{Dwyer_2012c,Hare_2016,Tran_2015,Wada_2019_prl}.
	In addition, a few cases of glow terminations coincided with downward TGFs \cite{Wada_2019_commphys,Smith_2018}.
	The connection between gamma-ray glows and downward TGFs is one of the intriguing questions in this field.

	One of the widely-known theories of electron acceleration in the atmosphere is the relativistic runaway electron avalanches (RREA) \cite{Gurevich_1992}.
	When the strength of electric fields exceeds the RREA threshold (0.284~MV~m$^{-1}$ at 1 atm \cite{Babich_2004,Dwyer_2003b}), 
	seed electrons of hundreds of keV or several MeV can be accelerated up to tens of MeV, and produce secondary electrons that are also accelerated.
	Therefore the number of electrons can be multiplied exponentially in such a high electric field.
	The accelerated electrons can emit bremsstrahlung photons as they interact with ambient atmospheric atoms.
	The energetic seed electrons are thought to be provided by cosmic rays, for example.
	The multiplication processes can be further boosted by the relativistic feedback processes \cite{Dwyer_2003b,Dwyer_2007}, 
	and some bright glows are though to require the feedback mechanism in fact \cite{Kelley_2015,Wada_2019_commphys}.
	
	Even when the electric fields are less than the RREA threshold, electrons of cosmic-ray origin can gain energies from the field while electron multiplication hardly takes place.
	The excess in energies from the normal cosmic-ray spectra can also be detected as fainter gamma-ray glows than those produced by RREA.
	This mechanism is called modification of the energy spectrum (MOS) \cite{Chilingarian_2012,Chilingarian_2014,Cramer_2017}, 
	and is the other important process to explain gamma-ray glows \cite{Ostgaard_2019,Bowers_2019}.

	In 2015, we launched a mapping observation campaign with multiple radiation monitors in the Hokuriku region \cite{Yuasa_2020}.
	After the launch of the campaign, we have reported the discovery of photonuclear reactions triggered by a downward TGF \cite{Enoto_2017}, 
	an observation of a glow termination with radio-frequency lightning mapping and electric field measurements \cite{Wada_2018}, 
	a simultaneous detection of a glow termination and a downward TGF \cite{Wada_2019_commphys}, and so on.
	Besides, more and more gamma-ray glows have been detected as the number of detectors increases.
	The present paper provides the first catalog of gamma-ray glows observed by the GROWTH collaboration during four winter seasons in Japan.
	Statistical investigations of occurence rates, temporary and spectral features of gamma rays are presented. 
	Throughout this paper, errors are at a confidence level of 1 standard deviation unless otherwise noted.

\section{Observation}

	Since 2006 we have operated an observation site at the Kashiwazaki-Kariwa nuclear power station of Tokyo Electric Power Company Holdings in Niigata Prefecture.
	In addition to the Kashiwazaki site, we established Kanazawa and Komatsu observation areas in Ishikawa Prefecture for the mapping observation campaign after 2015. 
	Their locations are presented in Figure~\ref{fig:map}.
	In the present paper we focus on the results obtained at the Kanazawa and Komatsu areas because very few gamma-ray glows have been detected at the Kashiwazaki site since 2015 
	despite several detections of downward TGFs and subsequent photonuclear reactions \cite{Enoto_2017,Wada_2019_prl}.

	The Kanazawa and Komatsu sites have been developed since 2016, after a preliminary observation campaign in the 2015-2016 winter season.
	In the four winter seasons from 2016-2017 (FY2016; FY stands for Japanese fiscal year, from April to next March) to 2019-2020 (FY2020), two observation sites in Komatsu and nine sites in Kanazawa were in operation.
	The location and operation period of the observation sites are summarized in Table~\ref{tab:detector}.
	Four sites are located at high schools, two at universities, two at public facilities, and the rest at a private company.
	Limited by the number of radiation detectors, up to 10 observation sites were simultaneously in operation (in FY2018 and FY2019).
	Figure~\ref{fig:operation} shows the time histories of the number of detectors in operation during four winter seasons.
	In each winter season, observations typically start in October or November, then end in March or April. 
	Therefore, the average operation duration is 4--5 months in each year.

	Our radiation monitors consist of two types of scintillation crystals; a $25.0 \times 8.0 \times 2.5~\mathrm{cm}^{3}$ bismuth germanate (Bi$_{4}$Ge$_{3}$O$_{12}$) or 
	a $30.0 \times 5.0 \times 5.0~\mathrm{cm}^{3}$ cesium iodide (CsI) crystal. While the BGO crystal is coupled with two Hamamatsu R1924A photomultipliers, the CsI crystal is with a Hamamatsu R6231 photomultiplier.
	Signals from the photomultipliers are amplified via a preamplifier and a shaping amplifier, then digitally sampled by an analog-to-digital convertor.
	Information of gamma rays, including energy and detection timing, is stored by Raspberry Pi. Details of the detection system are described in Yuasa et al. \cite{Yuasa_2020}.
	The energy range of radiation monitors varies because configurations (high voltage and lower threshold) can be changed in each year, but a typical range is 0.4--20.0~MeV. 	

	The absolute timing is conditioned by global positioning system (GPS) signals. When GPS signals are successfully received, the absolute timing accuracy is $\pm1~\mu$s.
	When signals are temporarily lost (during up to several hours), we interpolate the absolute timing with the GPS signals received before and after the lost. In this case the precision is better than 1~ms.
	When GPS signals are completely lost, the absolute timing is assigned by the internal clock of Raspberry Pi, conditioned by the network timing protocol (NTP) service. This case provides the timing precision of 1~second.
	If neither GPS signal nor internet connection is lost, we still rely on the internal clock of Raspberry Pi, but the absolute timing accuracy is unknown.
	In FY2016, the absolute accuracy is $\pm$0.5~sec even when GPS signals are successfully received due to a hardware issue. 
	This accuracy can be improved by further individual calibration as reported in Wada et al. \cite{Wada_2018}, 
	but we retain this accuracy in the present paper because this catalog analysis does not require more precision.
	Throughout this paper, we use Japan Standard Time (JST: UTC+9) to associate gamma-ray glows with daily variations of meteorological conditions in the local time.

\section{Results}
\subsection{Event detection}

	Gamma-ray glows can be identified as count-rate enhancements from background levels.
	The event scan has made by the method introduced in Yuasa et al. \cite{Yuasa_2020}.
	First, count-rate histories with 10-sec bins are produced. We employ the energy range above 3~MeV 
	because this range is not affected by radon-decay-chain background variations.
	With the 10-sec-binned count-rate histories, the standard deviation and the mean count rate are computed every 30 minutes.
	Then the significance of each bin is calculated; the count rate is divided by the standard deviation after the mean count rate is subtracted.
	If the significance of a bin exceeds the 5 standard deviations (5$\sigma$), it is considered as a potential glow event.
	Finally glow events are added to our event list by checking the count-rate histories manually.
	In this method, fainter events with a significance of less than 5$\sigma$ are usually ignored.
	On the other hand, glows with a significant of less that 5$\sigma$ can be coincidentally identified when searching over the data around the automatically-detected events.
	Such fainter glows are also added to the present event list.
	Most of the events were detected on thunder days in Kanazawa, confirmed by reports of Japan Meteorological Agency.
	The others were confirmed by a meteorological radar to coincide with a passage of a strong radar echo above the gamma-ray detector, 

	During the four winter seasons, in total 70 gamma-ray glows were detected. The event list is shown in Table~\ref{tab:event1}.
	Among the 70 gamma-ray glows, 66 glows have a significance of more than 5$\sigma$, and the remaining 4 glows have less significance.
	Twenty-four events have been reported in our previous publications \cite{Wada_2019_commphys,Yuasa_2020,Wada_2020,Wada_2021,Hisadomi_2021}.
	There are simultaneous detections with nearby detectors (events 2 and 3 for example). 
	They can be interpreted as an identical gamma-ray glow that passed over multiple radiation detectors with ambient wind flow \cite{Wada_2019_commphys,Yuasa_2020}.
	In the present paper, however, we consider a gamma-ray enhancement recorded by a radiation detector as an event in order to avoid interpretations in the event list. 

	Figure~\ref{fig:detection} shows the number of detected glow per day. The largest number of gamma-ray glows were detected on 17 February 2020 (8 events).
	The second largest number is 7 events on 10 January 2018 and 17 December 2018. 
	However, this data does not indicate the occurrence rates of gamma-ray glows
	because the number of deployed detectors differs as shown in Figure~\ref{fig:operation}.
	Figure~\ref{fig:ratio} presents the same data as Figure~\ref{fig:detection} but normalized by the number of radiation detectors in operation shown in Figure~\ref{fig:operation}.
	On 10 January 2018, 8 detectors were in operation and 7 events were detected. Hence in average 0.88 events were detected by a detector in the day.
	Nearly one event can be detected by one detector on days under heavy thunderstorms.
	The first peak of events typically comes in the beginning or middle of December. 
	Also there is sometimes the second peak in the middle of January.

\begin{figure}[tb]
	\begin{center}
	\includegraphics[width=0.9\hsize]{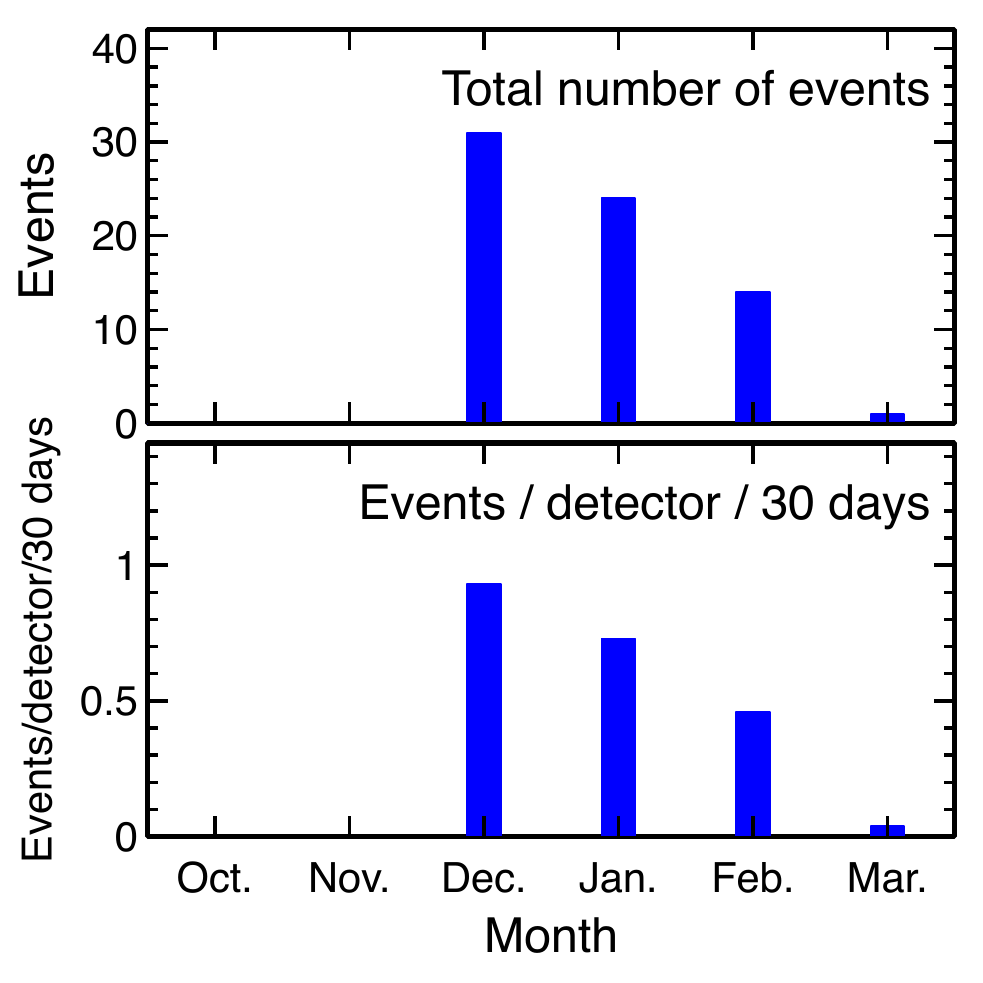}
	\caption{Figure~\ref{fig:par_month}: Monthly variations in the accumulated number of glow events. The upper panel shows the raw histogram, 
	and the lower shows the data normalized by the number of detectors in operation, hence indicates the averaged number of event detections by a radiation monitor in a month. }
	\label{fig:par_month}
	\end{center}
\end{figure}

	The monthly variation of glow detections is shown in Figure~\ref{fig:par_month}. 
	The upper panel presents accumulated data for 4 seasons.
	The largest number of glows were detected in December, while no events in October and November.
	This data should be also normalized by the number of detectors in operation as in the lower panel of Figure~\ref{fig:par_month}.
	December has still the largest number of glows, and most of the events were detected during December to February.
	In average every detector registered nearly one event in December.
	Though the detectors continue observation, glow events in March are quite rare (only event~70 on 16 March 2020).
	\textcolor{black}{A detailed discussion is given in the Discussion section.}
	
\begin{figure}[tb]
	\begin{center}
	\includegraphics[width=0.9\hsize]{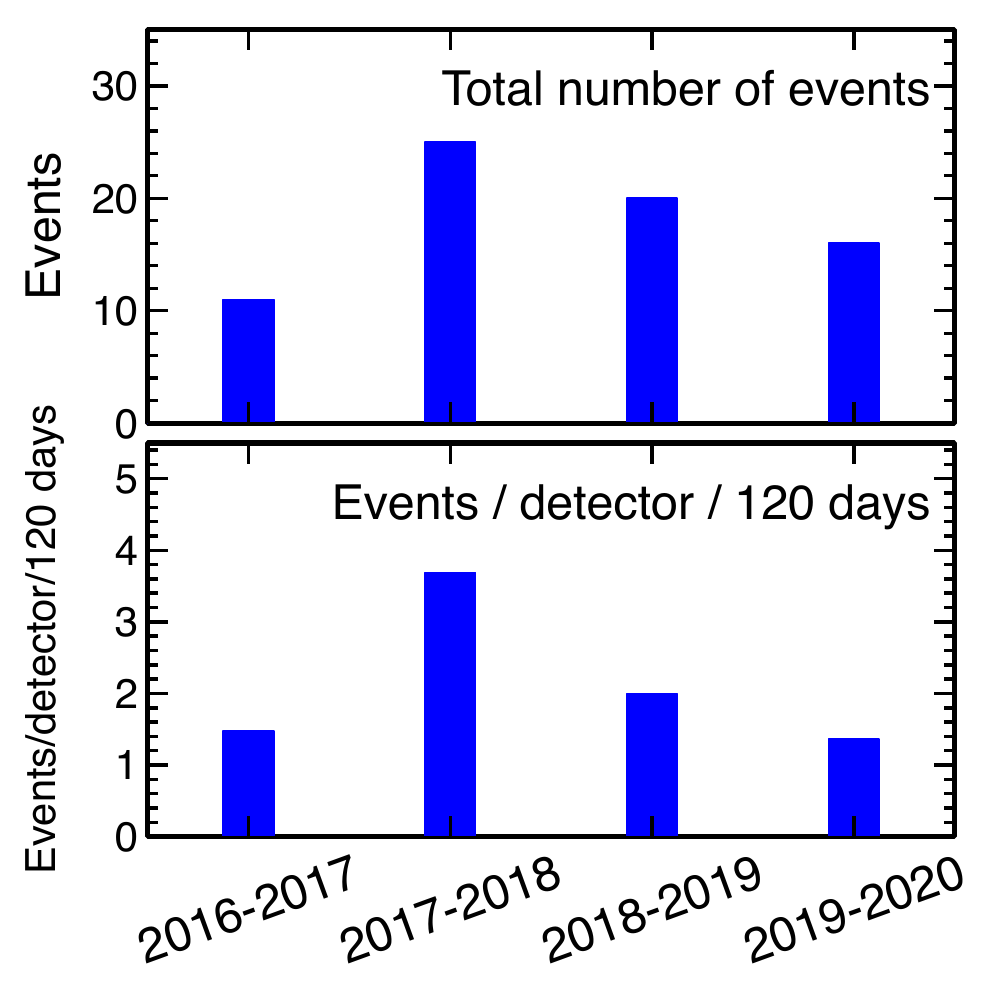}
	\caption{Figure~\ref{fig:par_year}: Yearly variations in the accumulated number of glow events. The upper panel shows the raw histogram, 
	and the lower shows the data normalized by the number of detectors in operation, hence indicates the averaged number of event detections by a radiation monitor in a winter seasons.
	Here a winter observation period is considered to continue 120 days (4 months) in average.}
	\label{fig:par_year}
	\end{center}
\end{figure}

	As well, the yearly variation of event detections is shown in Figure~\ref{fig:par_year}. In FY2017 (the 2017-2018 season), the largest number of events (25 glows) were detected.
	This trend is still true even after normalized by the number of deployed detectors as in the lower panel. 
	FY2017 was the most significant season for glow observations in four years; in average 3--4 events were detected by a detector during the season.
	The detection rates in the other three seasons were similar; 1--2 events by a detector during a season in average.

\begin{figure*}[tb]
	\begin{center}
	\includegraphics[width=0.8\hsize]{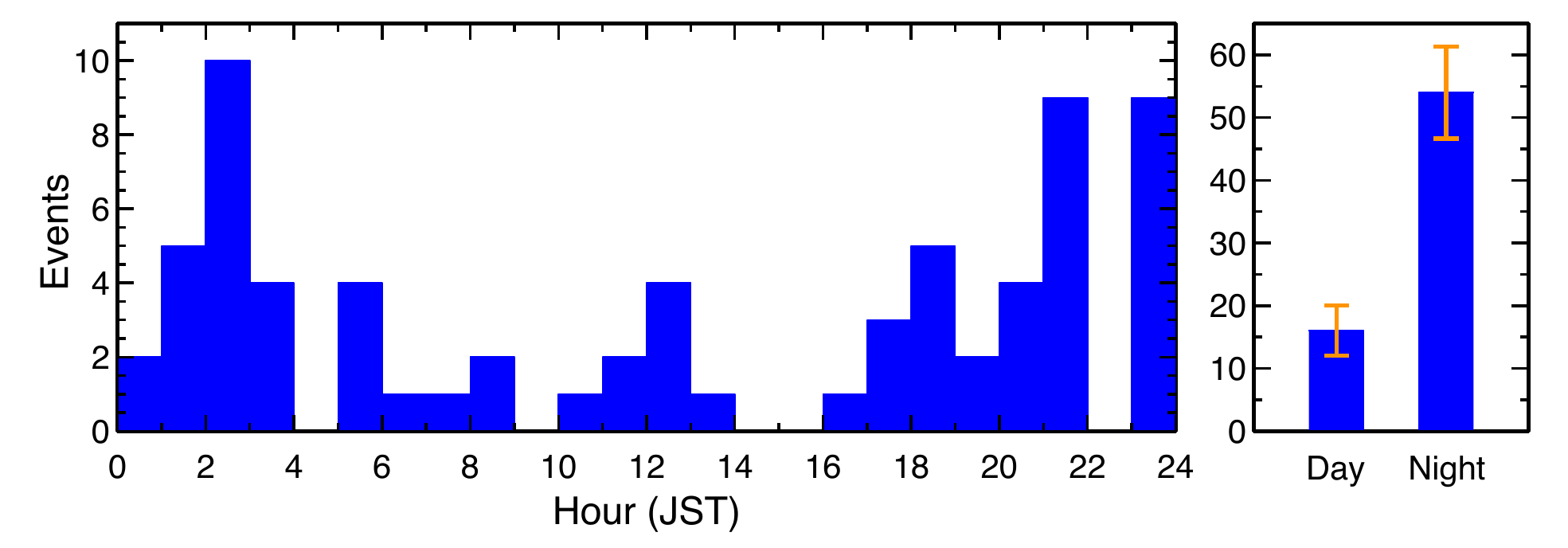}
	\caption{Figure~\ref{fig:par_hour}: Hourly variation in the accumulated number of glow events (left) and the difference in the number of detection (right)
	during daytime (06:00--18:00 JST) and nighttime (00:00--06:00 and 18:00--23:00 JST). The orange bars in the right panel indicate one standard deviation of the Poisson distribution.}
	\label{fig:par_hour}
	\end{center}
\end{figure*}

	The hourly variation of event detections are shown in Figure~\ref{fig:par_hour}. The largest number of events were detected during 02:00--03:00 JST.
	Also during 21:00--22:00 and 23:00--24:00 JST the second largest number of gamma-ray glows were registered. 
	While 23\% of events were detected during 06:00--18:00, 77\% were during 18:00-24:00 and 00:00--06:00.	
	Therefore nighttime tends to have more opportunities to detect gamma-ray glows than daytime.

\begin{figure*}[tb]
	\begin{center}
	\includegraphics[width=0.8\hsize]{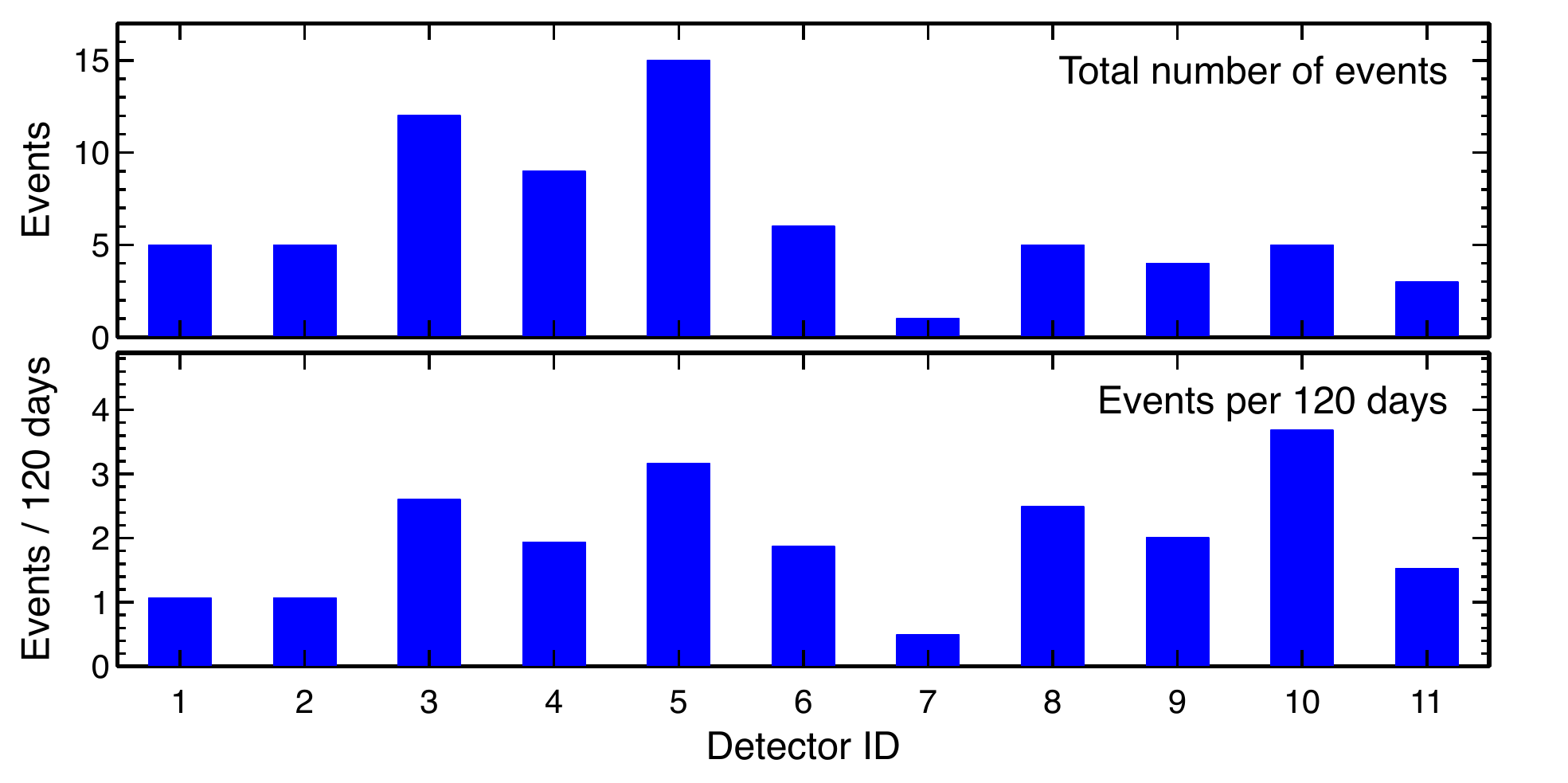}
	\caption{Figure~\ref{fig:par_detector}: The total number of detection (upper) and the normalized one (lower) for each observation site.
	The lower panel shows the detection frequency in 4~months, a typical observation period in a winter season.}
	\label{fig:par_detector}
	\end{center}
\end{figure*}

	The variations in the number of events by observation sites are shown in Figure~\ref{fig:par_detector}.
	As mentioned in Wada et al. \cite{Wada_2021}, the largest number of gamma-ray glows were detected at site 5, Kanazawa University Kakuma Campus; 15~events for four years.
	The lower panel shows the normalized distribution by operational days.
	The highest frequency of event detections is at site~10 (3.68~events/120~days). This site was operated in FY2018 and FY2019, but its operation was accidentally suspended in January in FY2019.
	Therefore the frequency is high for site~10 even though the total number of detection is only 5.

\subsection{Temporal analysis}

\begin{figure*}[tbh]
	\begin{center}
	\includegraphics[width=0.8\hsize]{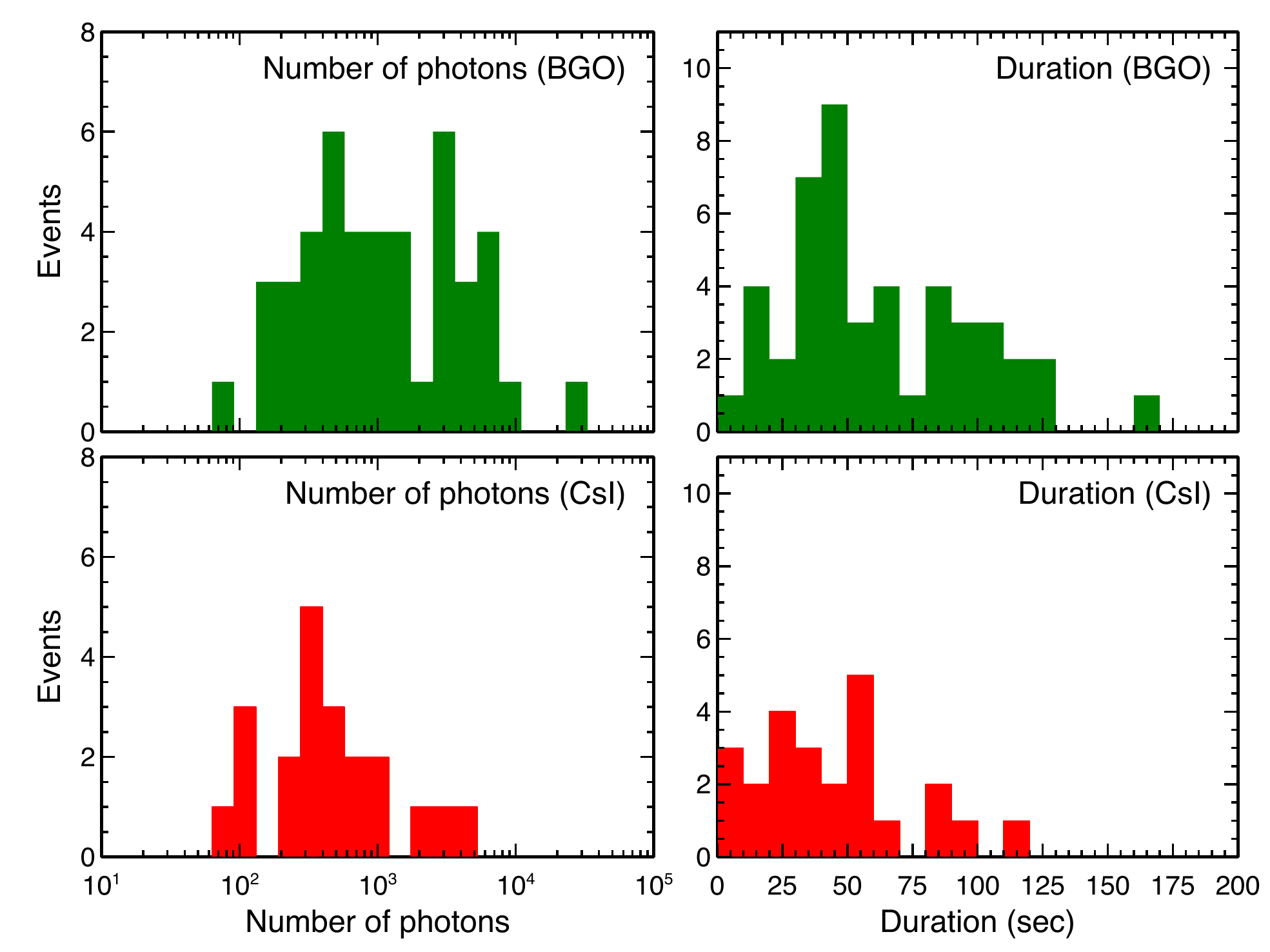}
	\caption{Figure~\ref{fig:photon-duration}: Histograms of the detected photon numbers (left) and the duration of gamma-ray glows registered with the BGO (upper) and CsI (lower) crystals.}
	\label{fig:photon-duration}
	\end{center}
\end{figure*}

	This subsection presents temporal analyses of individual events based on count-rate histories.
	Count-rate histories of all the 70 events are shown in Appendix~\ref{appendix}.
	First of all, the gamma-ray glows can be apparently classified into three types based on the shape of the histories;
	temporally-symmetric type with a single peak, temporally-asymmetric one with single or multiple peaks, and lightning-terminated one.
	The classification is included in Table~\ref{tab:event1}.
	Among the 70 detected glows, 34 events are the temporally-symmetric type, 17 are the temporally-asymmetric one, and the remaining 19 events are terminated with lightning discharges.

	The temporally-asymmetric type might be further divided into two types: single-peak and multiple-peak types.
	Clear examples of the multiple-peak type are Events~22 and 52. We can easily identify two peaks in an event.
	On the other hand, the boundary between the two types is vague. For example, Event~29 seems to have two peaks, but they are not well separated
	and hence can be also considered as a single-peaked asymmetric type.
	
	The lightning-terminated type has a sudden decrease in count rates. For most of the termination events except Event~38, 
	Japanese Lightning Detection Network (JLDN) reported one or multiple nearby lightning currents within 10~km from the observation sites 
	at the moment of termination, detected by the low-frequency measurement.
	Furthermore, the termination of Events~25, 58, and 59 coincided with a downward TGF. Events~25 and 58 were reported in Wada et al. \cite{Wada_2019_commphys}
	and Hisadomi et al. \cite{Hisadomi_2021}, respectively. The other cases coincided with no downward TGFs.

	To examine the brightness of the gamma-ray glows, the number of detected photons are calculated.
	To eliminate the background variations by radon-decay chain, here again signals above 3~MeV are counted.
	The method to constrain the number of photons is as follows; first a source window that covers the entire event is set, and photons above 3~MeV in the windows are counted up.
	Next, another window that have the same duration as the source window is set to a background period, and photons above 3~MeV in the windows are also counted up.
	The background period of each event is described in the public materials (see the Acknowledgment section).
	Then, the number of photons in the background window is subtracted from that in the source window.
	The uncertainty of this method is the difference in the photons in the background window and the actual background photons in the source window.
	Therefore, we consider the Poisson error of the photon number in the background window as the error of this estimation.
	This calculation is applied to the glows with $>5\sigma$ detection significance.
	For the glows terminated with a downward TGF, photons after the TGFs are not included to properly evaluate the gamma-ray glows.
		
	The results are shown in Table~\ref{tab:event1} and Figure~\ref{fig:photon-duration}.
	Due to the differences in detector responses / effective areas of the BGO and CsI crystals, the results are separately displayed.
	The largest number is 24081~photons and 4771~photons for detectors with BGO and CsI crystals, respectively.
	Its geometric mean is 1105~photons and 458~photons, and the median is 1024~photons and 383~photons, respectively.
	
	Another key parameter is duration. Here we define T80 duration; the duration window starts when the accumulated photon number exceeds 10\% of the total detected photon number calculated above, 
	and ends when the accumulated number exceeds 90\% of the total photon number. A similar method T90 is often used for gamma-ray bursts in astrophysics \cite{Zhang_2008}.
	The calculation was executed with 0.5-sec steps to consider deadtime correction.
	The results are listed in Table~\ref{tab:event1} and also displayed in Figure~\ref{fig:photon-duration}. 
	As well as the photon number, this calculation is applied to the glows with $>5\sigma$ detection significance, and the results from the BGO and CsI crystals are separated.
	The longest duration is 161.5~sec and 110.5~sec for detectors with BGO and CsI crystals, respectively.
	The average duration is 63.2~sec and 49.7~sec, and the median is 56.0~sec and 48.0~sec, respectively.
	When the results of BGO and CsI crystals are mixed up, the average duration is 58.9~sec.

\begin{figure}[t]
	\begin{center}
	\includegraphics[width=0.825\hsize]{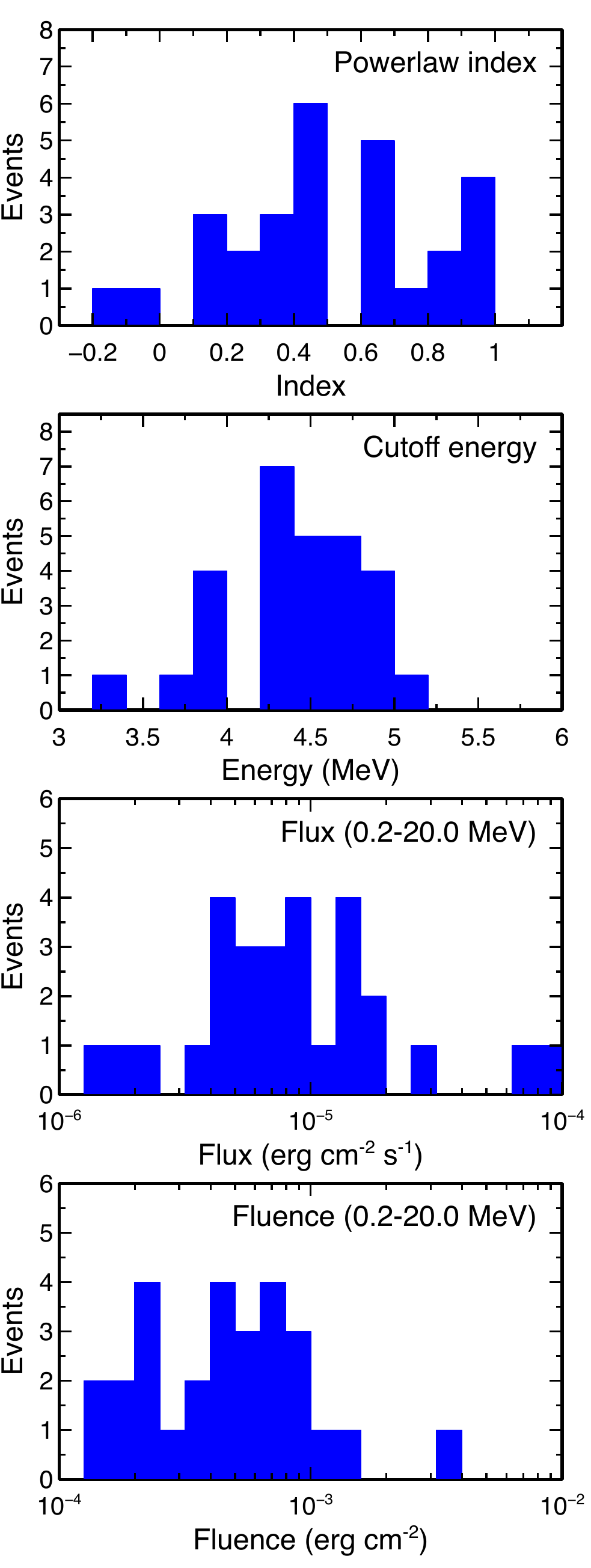}
	\caption{Figure~\ref{fig:spec_hist}: Histograms of fitting parameters obtained by spectral analysis.}
	\label{fig:spec_hist}
	\end{center}
\end{figure}

	Besides the calculation of the photon number and duration, we test a function fitting of the count-rate histories.
	Since the count-rate histories, in particular the temporally-symmetric type, may be approximated by a Gaussian function,
	Gaussian fitting to the histories are examined. The fitting function $F(t)$ is
	\begin{equation}\label{eq:gaussian}
		F(t) = C_{\rm bgd} + \sum_{i=1}^{n} a_{i} \exp \left\{ - \frac{(t - T_{i})^{2}}{2\sigma_{i}^{2}} \right\},
	\end{equation}
	where $t$ is time, $C_{\rm bgd}$ a constant value of the background count rate, $a$ the peak count rate, $T$ the peak time, $\sigma_{i}$ the standard deviation of the Gaussian function, 
	and $n$ the number of Gaussian functions to be added. $n$ should be as small as possible. 
	In most cases of the temporally-symmetric type, one Gaussian function can reproduce the count-rate histories, and hence $n = 1$.
	On the other hand, $n$ can be more than 1 for a part of the temporally-symmetric type that cannot be approximated by a Gaussian function, and for the temporally-asymmetric type.
	This fitting is also applied to a part of the lightning-terminated type if the events have a count-rate peak before the termination.
	In this case, a step function is folded to Equation~\ref{eq:gaussian} to imitate the termination. The moment of termination is set to be that of the lightning discharge reported by JLDN.
	It is noted that there is no physical reason to utilize the Gaussian function, but this is a tentative attempt to evaluate the shape of the count-rate histories.
	
	The fitting results are overlaid in Figure~\ref{fig:lightcurve1} to \ref{fig:lightcurve7}. 
	Most of the symmetric-type and a part of lightning-terminated events can be approximated by a single Gaussian function plus a background constant.
	A symmetric-type event (Event~26) and most of the asymmetric-type events can be approximated by two Gaussian functions plus a constant.
	Events~22 and 60 require three Gaussian components. The moment of termination in Event~15 is different from that reported by JLDN.
	Because the radiation monitor in Site~3 lost both GPS signals and the internet connection at that moment, the absolute timing is unreliable, 
	the discrepancy is not solid and thus we ignore it.  

\subsection{Spectral analysis}

\begin{figure}[tb]
	\begin{center}
	\includegraphics[width=\hsize]{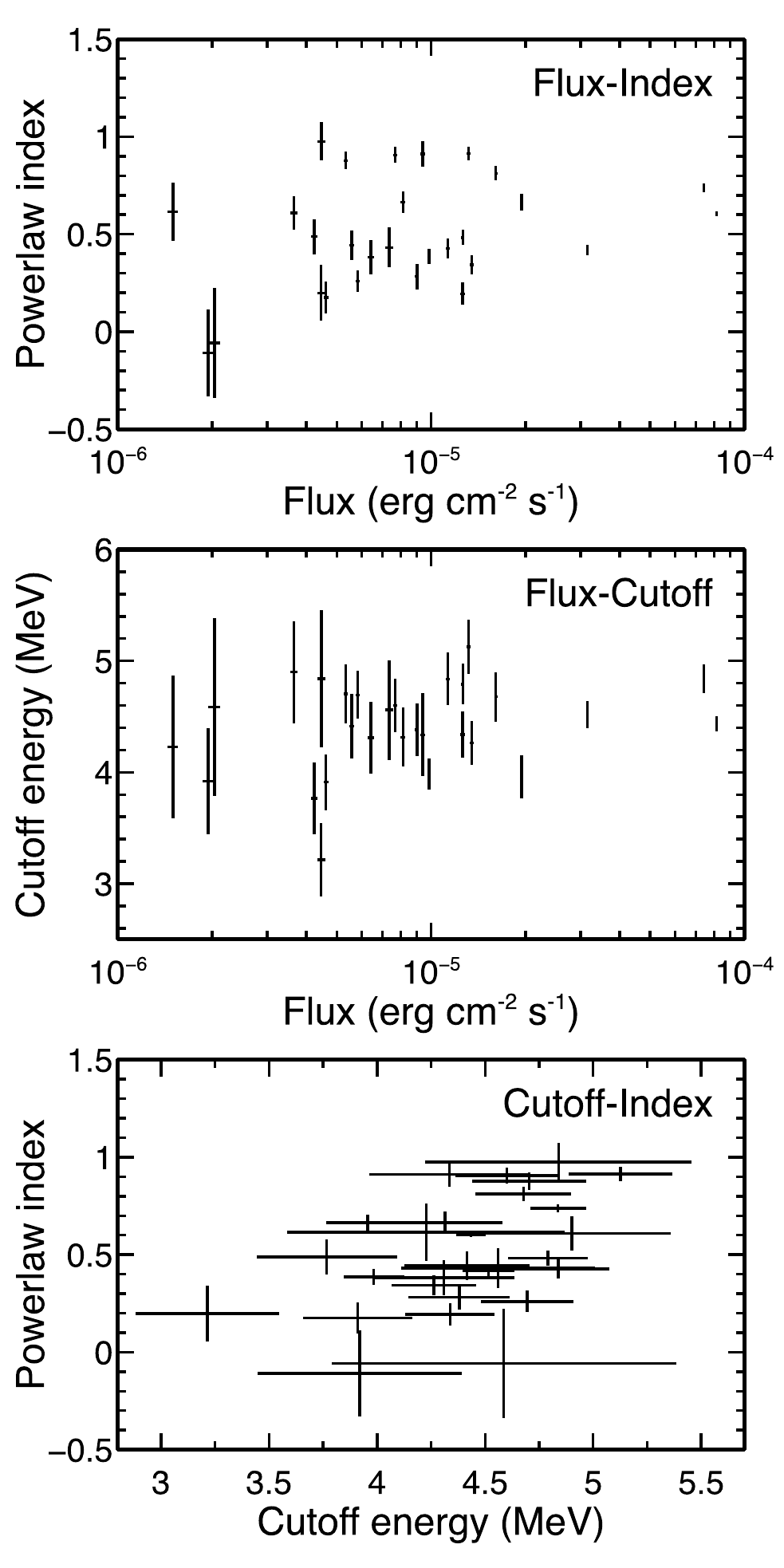}
	\caption{Figure~\ref{fig:spec_scatter}: Scatter plots of power-law index versus energy flux (upper), cutoff energy versus energy flux (middle), and power-law index versus cutoff energy (lower).}
	\label{fig:spec_scatter}
	\end{center}
\end{figure}

\begin{figure}[tb]
	\begin{center}
	\includegraphics[width=\hsize]{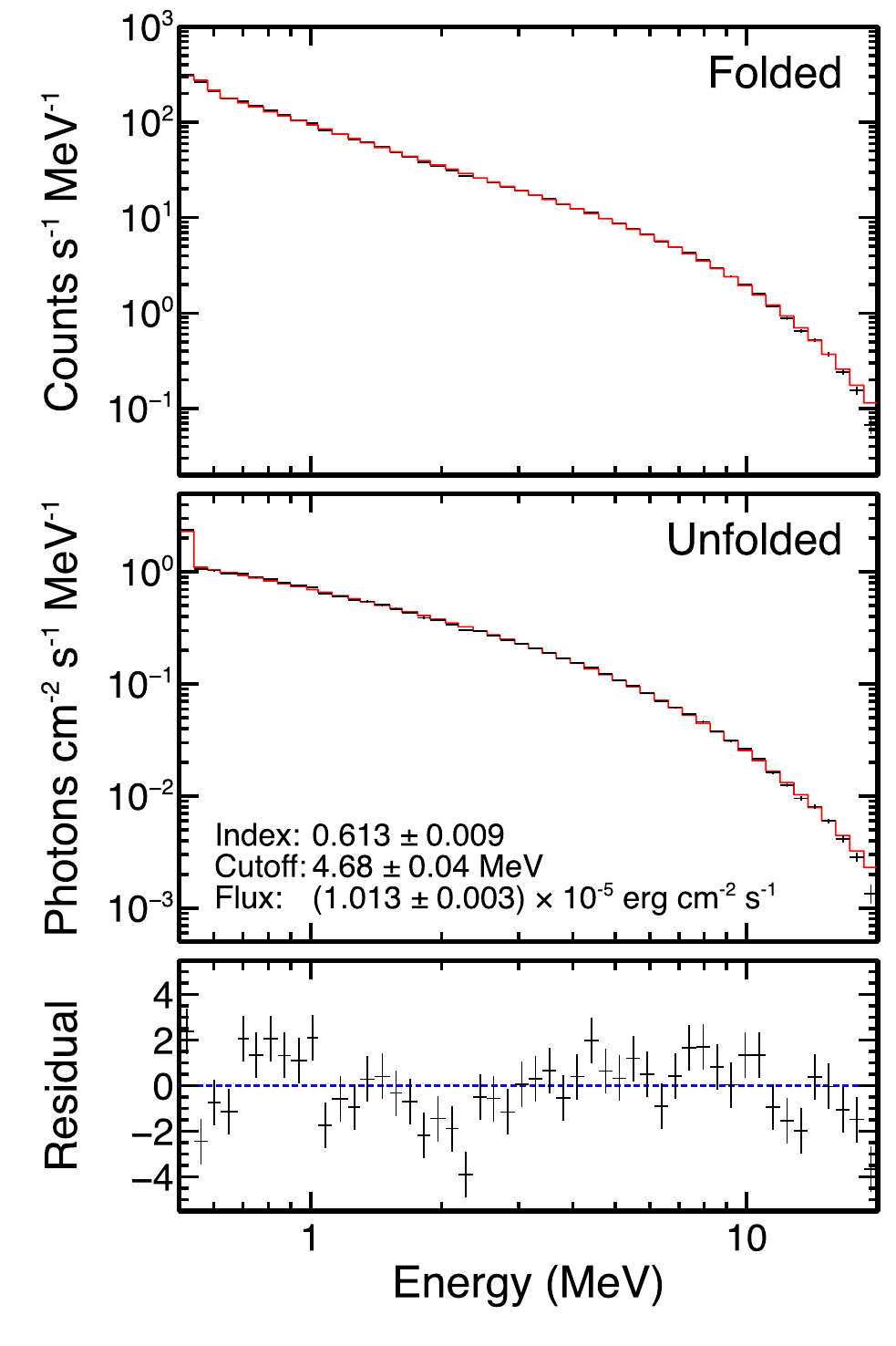}
	\caption{Figure~\ref{fig:spectrum_sum}: The averaged energy spectrum of gamma-ray glows detected by the BGO scintillators.
	The red lines show the best-fit function. Upper: the energy spectrum in which the detector responses are included. 
	Middle: the spectrum whose detector responses are unfolded by {\tt XSPEC} (i.e. the incident flux right before reaching the detector).
	Lower: the fitting residual; each data point of the spectrum is divided by its statistical error after the corresponding model value is subtracted.}
	\label{fig:spectrum_sum}
	\end{center}
\end{figure}

	We then examine spectral analysis of the gamma-ray glows. The analysis is performed with {\tt XSPEC}, a spectral analysis framework for high-energy astrophysics \cite{Arnaud_1996}.
	{\tt XSPEC} can unfold detector responses when an appropriate fitting function and the response matrix are given to derive the incident photon fluxes..
	The response matrices for the BGO and CsI scintillators were produced by Monte-Carlo simulations 
	with the simulation framework {\tt Geant4} \cite{Agostinelli_2003,Allison_2006,Allison_2016}, as shown in Yuasa et al. \cite{Yuasa_2020}.
	Among the 70 events, we focus on 28 events with $>$1000~photons above 3~MeV in order to avoid analyses with poor photon statistics.
	Also, a background spectrum is subtracted from the source spectrum. 
	Background spectra can change below 3~MeV as this energy band is affected by the radon-decay chain emission.
	It can lead a large systematic uncertainty of the background-subtracted spectra below 3~MeV if the intrinsic glow flux is low.
	Therefore, the spectral analysis is applied to the events with high photon statistics.
	
\begin{table*}[tbh]
\caption{Table~\ref{tab:spec}: The result of spectral fitting}
\begin{tabular}{c c c c c c}
\hline
No. & Crystal & Index & Cutoff & Flux$^{a}$ & Fluence$^{a}$ \\
    &         &       & (MeV)  & erg cm$^{-2}$~s$^{-1}$ & erg cm$^{-2}$ \\ \hline
3 & BGO & $0.17 \pm 0.08$ & $3.9^{+0.3}_{-0.2}$ & $(4.62 \pm 0.08) \times 10^{-6}$ & $(4.53 \pm 0.08) \times 10^{-4}$ \\
6 & BGO & $0.66 \pm 0.04$ & $3.96^{+0.2}_{-0.19}$ & $(1.94 \pm 0.02) \times 10^{-5}$ & $(8.16 \pm 0.1) \times 10^{-4}$ \\
13 & BGO & $0.66 \pm 0.06$ & $4.3 \pm 0.3$ & $(8.14 \pm 0.15) \times 10^{-6}$ & $(3.58 \pm 0.06) \times 10^{-4}$ \\
14 & BGO & $0.62^{+0.14}_{-0.15}$ & $4.2^{+0.7}_{-0.6}$ & $(1.5 \pm 0.06) \times 10^{-6}$ & $(1.62 \pm 0.06) \times 10^{-4}$ \\
16 & BGO & $0.91^{+0.06}_{-0.07}$ & $4.3 \pm 0.4$ & $(9.39 \pm 0.18) \times 10^{-6}$ & $(2.4 \pm 0.05) \times 10^{-4}$ \\
19 & BGO & $0.19 \pm 0.06$ & $4.3 \pm 0.2$ & $(1.26 \pm 0.02) \times 10^{-5}$ & $(4.35 \pm 0.07) \times 10^{-4}$ \\
20 & BGO & $0.43 \pm 0.05$ & $4.8 \pm 0.2$ & $(1.13 \pm 0.018) \times 10^{-5}$ & $(4.41 \pm 0.07) \times 10^{-4}$ \\
22 & BGO & $0.39 \pm 0.04$ & $3.98 \pm 0.14$ & $(9.84 \pm 0.11) \times 10^{-6}$ & $(9.06 \pm 0.1) \times 10^{-4}$ \\
24 & BGO & $0.81 \pm 0.04$ & $4.7 \pm 0.2$ & $(1.61 \pm 0.02) \times 10^{-5}$ & $(5.71 \pm 0.07) \times 10^{-4}$ \\
25 & BGO & $0.74 \pm 0.02$ & $4.84 \pm 0.13$ & $(7.42 \pm 0.05) \times 10^{-5}$ & $(1.336 \pm 0.01) \times 10^{-3}$ \\
26 & BGO & $0.604 \pm 0.013$ & $4.43 \pm 0.07$ & $(8.15 \pm 0.04) \times 10^{-5}$ & $(3.83 \pm 0.017) \times 10^{-3}$ \\
29 & CsI & $0.2^{+0.13}_{-0.15}$ & $3.2 \pm 0.3$ & $(4.46 \pm 0.14) \times 10^{-6}$ & $(2.14 \pm 0.06) \times 10^{-4}$ \\
30 & CsI & $0.98 \pm 0.1$ & $4.8^{+0.7}_{-0.6}$ & $(4.47 \pm 0.13) \times 10^{-6}$ & $(2.37 \pm 0.07) \times 10^{-4}$ \\
43 & BGO & $-0.1 \pm 0.3$ & $4.6^{+0.9}_{-0.7}$ & $(2.04 \pm 0.09) \times 10^{-6}$ & $(1.28 \pm 0.06) \times 10^{-4}$ \\
44 & BGO & $-0.1 \pm 0.2$ & $3.9^{+0.5}_{-0.4}$ & $(1.94 \pm 0.07) \times 10^{-6}$ & $(1.74 \pm 0.06) \times 10^{-4}$ \\
46 & CsI & $0.38 \pm 0.09$ & $4.3 \pm 0.3$ & $(6.43 \pm 0.14) \times 10^{-6}$ & $(3.63 \pm 0.08) \times 10^{-4}$ \\
50 & BGO & $0.42 \pm 0.03$ & $4.52 \pm 0.12$ & $(3.15 \pm 0.03) \times 10^{-5}$ & $(1.009 \pm 0.009) \times 10^{-3}$ \\
52 & BGO & $0.61^{+0.08}_{-0.09}$ & $4.9^{+0.5}_{-0.4}$ & $(3.65 \pm 0.09) \times 10^{-6}$ & $(2.52 \pm 0.06) \times 10^{-4}$ \\
55 & BGO & $0.43^{+0.1}_{-0.11}$ & $4.6^{+0.5}_{-0.4}$ & $(7.3 \pm 0.2) \times 10^{-6}$ & $(1.43 \pm 0.04) \times 10^{-4}$ \\
57 & BGO & $0.34 \pm 0.05$ & $4.26^{+0.2}_{-0.19}$ & $(1.35 \pm 0.02) \times 10^{-5}$ & $(4.58 \pm 0.07) \times 10^{-4}$ \\
58 & BGO & $0.48 \pm 0.04$ & $4.79^{+0.19}_{-0.18}$ & $(1.263 \pm 0.015) \times 10^{-5}$ & $(7.45 \pm 0.09) \times 10^{-4}$ \\
60 & BGO & $0.88^{+0.04}_{-0.05}$ & $4.7 \pm 0.3$ & $(5.35 \pm 0.08) \times 10^{-6}$ & $(5.94 \pm 0.09) \times 10^{-4}$ \\
65 & BGO & $0.91 \pm 0.04$ & $4.6 \pm 0.2$ & $(7.67 \pm 0.1) \times 10^{-6}$ & $(6.6 \pm 0.08) \times 10^{-4}$ \\
66 & BGO & $0.91 \pm 0.04$ & $5.1 \pm 0.2$ & $(1.317 \pm 0.016) \times 10^{-5}$ & $(6.52 \pm 0.08) \times 10^{-4}$ \\
67 & CsI & $0.28^{+0.06}_{-0.07}$ & $4.4 \pm 0.2$ & $(9.02 \pm 0.13) \times 10^{-6}$ & $(8.08 \pm 0.12) \times 10^{-4}$ \\
68 & CsI & $0.44^{+0.07}_{-0.08}$ & $4.4 \pm 0.3$ & $(5.58 \pm 0.1) \times 10^{-6}$ & $(6.17 \pm 0.11) \times 10^{-4}$ \\
69 & BGO & $0.26^{+0.05}_{-0.06}$ & $4.7 \pm 0.2$ & $(5.82 \pm 0.08) \times 10^{-6}$ & $(6.99 \pm 0.1) \times 10^{-4}$ \\
70 & BGO & $0.49 \pm 0.09$ & $3.8 \pm 0.3$ & $(4.23 \pm 0.1) \times 10^{-6}$ & $(2.45 \pm 0.06) \times 10^{-4}$ \\
\hline
(average) & BGO & 0.613 $\pm$ 0.009 & 4.68 $\pm$ 0.04  & $(1.013 \pm 0.003) \times 10^{-5}$ & -- \\
\hline
\multicolumn{6}{l}{$^{a}$ The energy range is 0.2--20.0~MeV.}\\
\end{tabular}
\label{tab:spec}
\end{table*}	
	
	The source spectrum for each event is extracted from the T80 window, namely the window between two blue-dashed lines in Figure~\ref{fig:lightcurve1} to \ref{fig:lightcurve7} of Appendix~\ref{appendix}.
	The window for the background spectrum is set to have the same duration as the source window, not to include the gamma-ray glow, but to include a nearby background period. 
	The background-subtracted spectra are fitted by a power-law function with an exponential cutoff $S(E)$
	\begin{equation}
		S(E) = A \times E^{-\Gamma} \exp \left(-\frac{E}{E_{\rm cut}} \right),
	\end{equation}
	where $E$, $A$, $\Gamma$, and $E_{\rm cut}$ are energy of gamma-ray photons, normalization factor, power-law index, and cutoff energy, respectively.
	This is an approximated form of a bremsstrahlung spectrum produced by RREA electrons \cite{Dwyer_2012b}.
	The energy flux in 0.2--20.0~MeV $F_{0.2-20}$ is derived as
	\begin{equation}
	F_{0.2-20} = \int^{\rm 20~MeV}_{\rm 0.2~MeV} A \times E^{-\Gamma} \exp \left(-\frac{E}{E_{\rm cut}} \right) EdE
	\end{equation}
	by extrapolating the gamma-ray spectrum down to 0.2~MeV with best-fit parameters.
	The fitting energy range varies by events because the energy range depends on detectors, due to differences in amplifier gain, lower and upper thresholds.
	The fitting is performed to cover the full energy range of each event, but uses above 0.6~MeV for all the events 
	because the range below 0.6~MeV can be affected by the annihilation line at 0.511~MeV and hence not be fitted only with the power-law function.

\begin{figure*}[tbh]
	\begin{center}
	\includegraphics[width=\hsize]{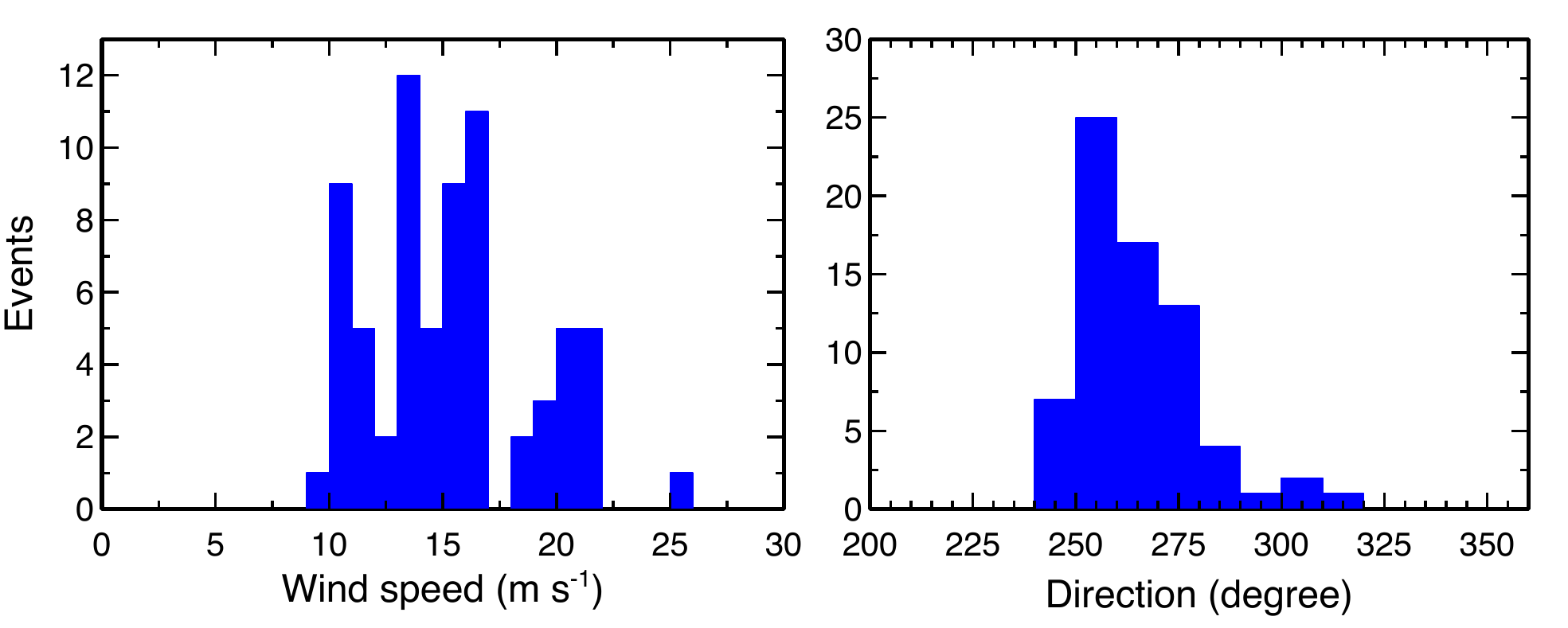}
	\caption{Figure~\ref{fig:weather}: Histograms of wind speed and its direction at the moment of glow detections. 
	The direction is where the wind comes from in clockwise. For example, a direction of 270$^{\circ}$ means a wind comes from west toward east.}
	\label{fig:weather}
	\end{center}
\end{figure*}

\begin{figure}[tbh]
	\begin{center}
	\includegraphics[width=\hsize]{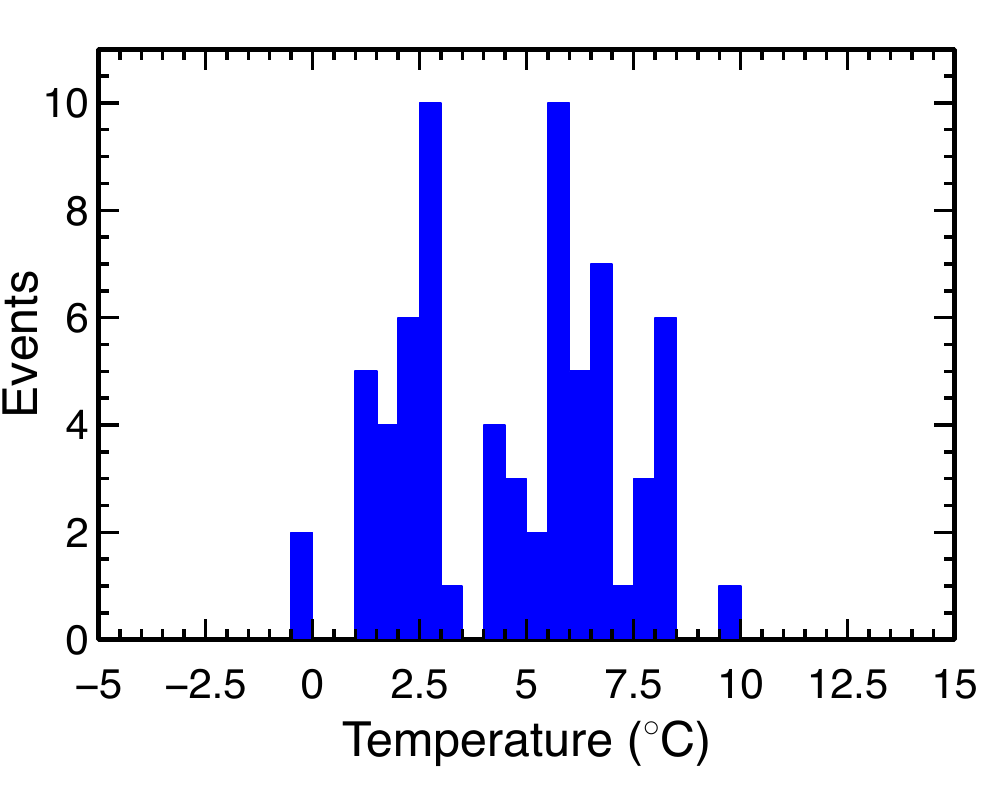}
	\caption{Figure~\ref{fig:temp}: A histogram of atmospheric temperature during gamma-ray glows, measured by Japan Meteorological Agency.}
	\label{fig:temp}
	\end{center}
\end{figure}

	All the response-folded and response-unfolded spectra are shown in Figure~\ref{fig:spectrum1} to Figure~\ref{fig:spectrum6}. The best-fit functions are also overlaid.
	The fitting parameters are listed in Table~\ref{tab:spec}. Histograms of the parameters are also shown in Figure~\ref{fig:spec_hist}.
	The power-law index is 0.50 on average, with a standard deviation of 0.28. As well the average of the cutoff energy is 4.41~MeV, with a standard deviation of 0.41~MeV.
	The energy flux varies from $1.5\times10^{-6}$ to $8.1\times10^{-5}$~erg~cm$^{-2}$~s$^{-1}$, with the geometric mean of $8.4\times10^{-6}$~erg~cm$^{-2}$~s$^{-1}$.
	Scatter plots between two parameters of three are shown in Figure~\ref{fig:spec_scatter}.
	There is no significant correlations between the energy flux and the power-law index nor between the energy flux and the cutoff energy.
	This feature is different from observations in Aragats \cite{Chilingarian_2014}, which reports a clear positive correlation between power-law index and intensity at 1~MeV. 
	The power-law index and the cutoff energy have a moderate negative correlation with a correlation coefficient of 0.47.
	
	We then extract an averaged spectrum to derive typical parameters of glow spectra. 
	Here the energy spectra of 21 glows detected by the BGO crystals are accumulated 
	because the spectra of BGO and CsI crystals cannot be summed up due to the difference in detector responses.
	Also Events~3 and 6 were excluded from this analysis because the energy range of these events are narrower than the others, and hence they would reduce the quality of the analysis.
	The fitting range is set to 0.5--20.0~MeV. For spectral fitting, a Gaussian function is added to the power-law function in order to reproduce the annihilation line at 0.511~MeV.
	The total exposure is 1234.0~sec. The averaged spectrum and the fitting result are shown in Figure~\ref{fig:spectrum_sum}. The reduced $\chi^{2}$ of the best-fit function is 2.3 with 46 degrees of freedom.
	Even after unfolding the detector responses, the exponential roll-off can be clearly seen around 5~MeV.
	The power-law index, cutoff energy, and energy flux of the best-fit function are $0.613\pm0.009$, $4.68\pm0.04$~MeV, 
	and $(1.013\pm0.003)\times10^{-5}$~erg~cm$^{-2}$~s$^{-1}$ (0.2-20.0~MeV), respectively.
	The annihilation line has an equivalent width of $38.8^{+1.7}_{-2.5}$~keV.

\subsection{Meteorological conditions}

	Meteorological conditions, including ambient wind and atmospheric temperature are of great interest for investigation of gamma-ray glows.
	The wind speed and its direction at the moment of glow detections can be derived from meteorological radar data, 
	by calculating moving vector with two time-separated radar contours. The detail method of the calculation is introduced in Wada et al. \cite{Wada_2019_commphys}.
	This method works under the assumption that a cloud or an echo region moves with ambient wind.
	We utilize the data of the eXtended RAdar Information Network (XRAIN), operated by Ministry of Land, Infrastructure, Transport and Tourism of Japan.
	Both Komatsu and Kanazawa areas are covered by an X-band (9.8~GHz) dual-polarized radar at the Noumi radar site of XRAIN.
	Synthetic precipitation contours or plan-position indicator contours of reflectivity factor obtained every minute are used for the wind estimation.
	
	The estimated wind velocity and direction are listed in Table~\ref{tab:event1}. Also their histograms are shown in Figure~\ref{fig:weather}.
	The direction is where the wind comes from in clockwise. For example, a direction of 270$^{\circ}$ means a wind comes from west toward east.
	The average velocity is 15.3~m~s$^{-1}$ with a standard deviation of 3.5~m~s$^{-1}$, and the average direction is 260$^{\circ}$ with a standard deviation of 15$^{\circ}$.
	Most of the gamma-ray glows occurred during west or west-southwest winds, and a few during west-northwest winds.
	Gamma-ray glows never occurred during north, south, nor east winds. 
	
	Atmospheric temperature during glows and its histogram are also presented in Table~\ref{tab:event1} and Figure~\ref{fig:temp}, respectively.
	The temperature was measured at Komatsu AMeDAS Station (N36.382$^{\circ}$, E136.436$^{\circ}$; for the events at sites~1 and 2) 
	and Kanazawa Meteorological Observatory (N36.589$^{\circ}$, E136.634$^{\circ}$; for the other events) of Japan Meteorological Agency.
	The histories of the temperature measurement can be retrieved from the web site of Japan Meteorological Agency.
	The average temperature is 4.6$^{\circ}$C with a standard deviation of 2.4$^{\circ}$C. Most cases occurred at a temperature above 0$^{\circ}$C,
	but two cases were at below 0$^{\circ}$C (Events~7 and 30).
	
\section{Discussion}

\begin{figure*}[tbh]
	\begin{center}
	\includegraphics[width=\hsize]{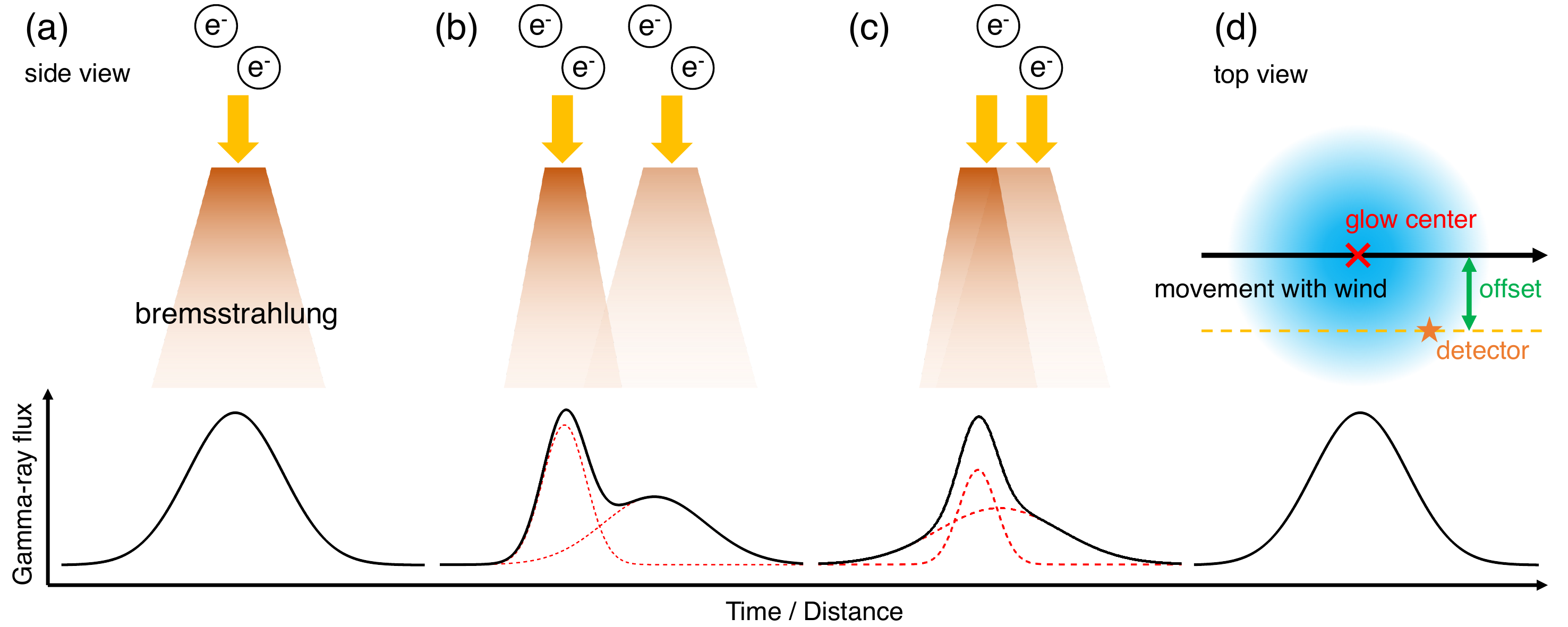}
	\caption{Figure~\ref{fig:schematics}: Side-viewed schematics of the temporally-symmetric~(a), double-peaked~(b), temporally-asymmetric~(c) types, 
	and a top-viewed relation between a temporally-symmetric glow and a radiation detector~(d). The blue region in panel~(d) shows the radiation area of a gamma-ray glow on the ground.}
	\label{fig:schematics}
	\end{center}
\end{figure*}

\begin{figure}[tb]
	\begin{center}
	\includegraphics[width=\hsize]{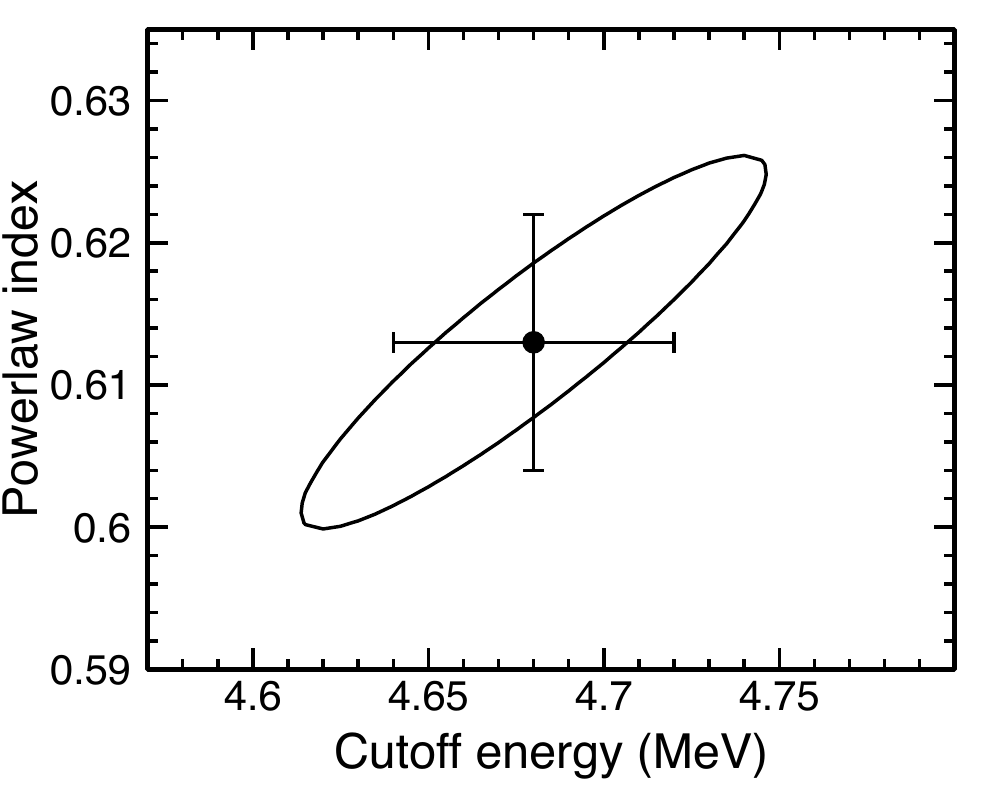}
	\caption{Figure~\ref{fig:fitting_correlation}: A confidence region at one standard deviation of a fitting for the averaged spectrum, in the space of the power-law index and the cutoff energy.
	The cross marker also shows one standard deviation errors of the two parameters but independently calculated.}
	\label{fig:fitting_correlation}
	\end{center}
\end{figure}

	Almost all the glows were detected in December to February, except one events in March 2020.
	In particular, the detection frequency is the highest in December as shown in the lower panel of Figure~\ref{fig:par_month}.
	Winter thunderstorms in the Hokuriku region usually starts at the end of November or the beginning of December, and ends at the beginning of March.
	According to the records by Japan Meteorological Agency, the averaged thunder days in Kanazawa during the four winter seasons (FY2016 to FY2019) are 
	2.3, 5.5, 6.0, 3.0, and 2.8 days in November, December, January, February, and March, respectively.
	Therefore, the monthly variation of glow detections roughly corresponds to that of thunder days, even though the relation between them is not necessarily proportional.
	The thunder days in December to February are 17, 18, 10, and 10 days in FY2016, FY2017, FY2018, and FY2019, respectively.
	It is also consistent that FY2017, in which the largest number of glows were detected, had the largest number of thunder days.
	We note that during 1981--2020, the averaged thunder days in three months from December to February are 21.1 days (according to Japan Meteorological Agency),
	significantly higher than in the four recent years presented here.

	\textcolor{black}{As in the Introduction section, a typical cloud base of winter thunderclouds is several hundreds meters, significantly lower than summer ones, 
	and a typical cloud-top altitude is 5--6~km. \cite{Goto_Narita_1992,Wada_2018,Wada_2021}
	Since the charge center is close to the ground, gamma rays can reach detectors at sea level before they are attenuated in the atmosphere.
	The average temperature at the moment of glow detections is 4.6$^{\circ}$C with a standard deviation of 2.4$^{\circ}$C, 
	which is significantly lower than one when summer thunderstorms occur. Also the westerly wind is predominant in most cases.
	In the Hokuriku region, including Kanazawa and Komatsu, the north wind typically causes cold air to flow in, causing the surface temperature to drop below freezing and cause snowfall.
	At this time, since the altitude at 0$^{\circ}$C where charging occurs reaches the ground, lightning does not occur in many cases, with some exceptions.
	Maintaining a surface temperature of around 5$^{\circ}$C with the westerly wind is a typical meteorological condition for gamma-ray glow.
	We note that it is necessary to discuss whether the conditions under which winter thunderstorms occur and ones under which gamma-ray glows occur are the same or different. 
	A further study of winter thunderstorms themselves is needed to answer this question.}
	
	As described in the Results section, the majority of the gamma-ray glows has a symmetric count-rate history that can be approximated by one Gaussian function.
	This is suggestive of the idea that the irradiation area has a circle or ellipse shape, and passes above a radiation detector with a constant wind flow.
	A schematics of this idea is shown in Figure~\ref{fig:schematics}. In fact, Events~2 and 3 can be explained by a passage of a thundercloud, presented in Yuasa et al. \cite{Yuasa_2020}
	For the temporally-asymmetric type, there are several interpretations. One is that multiple electron-acceleration regions adjoin each other, as in Figure~\ref{fig:schematics}.
	This idea is supported by the result that the temporally-asymmetric glows are approximated by two or three Gaussian functions.
	In fact, several temporally-asymmetric glows have two peaks clearly (Events~22, 52, and 60) and they can be explained by this idea.
	Another one is that the irradiation area does not have a circle nor ellipse shape, but a more complex shape. 
	This can be caused by a complex shape or a tilt of electron-acceleration region.
	It is also possible that an avalanche region rapidly glows and decays \cite{Hisadomi_2021,Chilingarian_2020b}.
	It is difficult to distinguish these ideas for the temporally-asymmetric type with a single peak.
	As mentioned before, using Gaussian functions is tentative and empirical just to evaluate the shape of the gamma-ray glows, without enough physical reasoning. 
	Therefore, some temporally-asymmetric events could be explained and evaluated by other functions.
	
	The spectral fittings allow us to reconfirm that the energy spectra of gamma-ray glows with high photon statistics can be approximated by the power-law function with an exponential cutoff,
	and hence can be explained by bremsstrahlung emission from electrons.
	The average cutoff energy, 4.41~MeV with a standard deviation of 0.41~MeV, is close to the average energy of RREA electrons ($\sim$7.3~MeV \cite{Dwyer_2012b}). 
	As in Fig.~3 of Dwyer et al. \cite{Dwyer_2012b}, the average energy of RREA electrons with an electric field below 0.5~MV~m$^{-1}$ can be lower than 7.3~MeV.
	Since gamma-ray glows are considered to occur in an electric field close to the RREA threshold, being different from TGFs, 
	the obtained cutoff energy might reflect the electric field strength inside thunderclouds. However, the spectral shape is also affected by atmospheric scattering and absorption,
	and hence we need further investigations considering atmospheric propagation of gamma rays with Monte-Carlo simulations.
	
	As seen in the lower panel of Figure~\ref{fig:spec_scatter}, there is a positive correlation between power-law index and cutoff energy.
	However, these two parameters are not independent. Figure~\ref{fig:fitting_correlation} shows 
	a confidence area at one standard deviation in the space of the power-law index and the cutoff energy, derived from the averaged energy spectrum.
	As the confidence area is a narrow ellipse, the parameters are coupled. 
	It means that a function with a larger index and a larger cutoff energy is similar to that with a smaller index and a smaller cutoff energy.	
	Therefore, this positive correlations is caused by the characteristics of the fitting function, and do not necessarily have physical meanings.

	Chilingarian et al. \cite{Chilingarian_2019b} published a catalog of gamma-ray glows/TGEs detected at Mt. Aragats in Armenia in 2017.
	It is valuable to compare it with the present work. In 2017, 44 TGEs with high-energy particles (above 3~MeV) were detected at the Aragats Space Environment Center (ASEC).
	One of the significant differences is the seasons. They are all connected to summer thunderstorms from April to November while our cases were during winter thunderstorms.
	Since ASEC is located at an altitude of 3200~m, a typical temperature at the moment of TGE detections ranges from $-3^{\circ}$C to $3^{\circ}$C even in summer, 
	lower than the average temperature of our cases.
	
	Another significant difference is the number of detections; 44 TGEs from April to November, and hence 8.8 TGEs per month at a single site in their case.
	In the present analysis, the detection frequency is as many as 0.93~events per month per observation site.
	Therefore, TGEs/gamma-ray glows were detected more frequently at ASEC than in the Hokuriku region.
	This is potentially caused by the difference in meteorological conditions, namely the frequency of lightning discharges at Aragats and Hokuriku.
	Besides that, there is also a difference in atmospheric temperature. At ASEC, TGEs were detected frequently even below 0$^{\circ}$.
	In these cases, the cloud base can get closer to the ground, and it is possible that sometimes the detectors are inside thunderclouds.
	This indicates that the detectors at ASEC is often closer to the acceleration region than our cases, 
	and they could have more opportunities to detect TGEs due to less atmospheric absorption of gamma rays.
	
\section{Conclusion}

	Thanks to the mapping observation campaign with up to 10 radiation monitors for four winter seasons, 
	we observed 70 gamma-ray glows during winter thunderstorms in Japan, which allows us a statistical analysis.
	Detections of almost all the glows concentrated on December, January, and February, in which winter thunderstorms are frequent in the Hokuriku region, and 77\% were detected in nighttime. 
	A typical T80 duration is minute-scale, 58.9~sec on average. A half of the detected glows has a temporally-symmetric (Gaussian-like) count-rate variation, 
	suggestive of a stable electron accelerator passing above the detectors with a constant wind flow. 
	A quarter has a temporally-asymmetric variation, implying multiple acceleration regions or one region with a complex shape. The rest quarter terminated with lightning discharges.
	Energy spectra are well approximated by the power-law function with an exponential cutoff, whose average power-law index, cutoff energy, and energy flux are
	0.50, 4.41~MeV, and $8.4\times10^{-6}$~erg~cm$^{-2}$~s$^{-1}$, respectively. During the glow detections, winds flow typically eastward, and the temperature is 4.6$^{\circ}$C on average.
	It will be valuable if the enriched data of the present catalog is further investigated with other observational, theoretical and simulation works.

\begin{acknowledgments}

	We deeply thank providers of the observation sites; K.~Watarai and Kanazawa University High School, K.~Yoneguchi and Kanazawa Izumigaoka High School, 
	K.~Kimura and Komatsu High School,  K.~Kitano and Science Hills Komatsu, K. Kono and Ishikawa Plating Industry Co., Ltd,
	S.~Kura and Kanazawa Nishi High School, T.~Matsui and Industrial Research Institute of Ishikawa.
	The BGO scintillation crystals are provided by Nuclear Experimental Group, Graduate School of Science, The University of Tokyo, thanks to Dr. H.~Sakurai and Dr. M.~Niikura.
	This project is supported by JSPS/MEXT KAKENHI grants 15K05115, 16H06006, 18J13355, 18H01236, 19H00683, 20K22354, 21H01116, 
	the Hakubi project and SPIRITS 2017 of Kyoto University, the joint research program of the Institute for Cosmic Ray Research (ICRR), The University of Tokyo,
	RIKEN Special Postdoctoral Researchers program, and RIKEN Hakubi Research Fellowship program. 
	The bootstrapping phase of this project was supported by a crowdfunding campaign named Thundercloud Project on the academic crowdfunding platform “academist.”
	The observation data and materials to produces figures are available on Mendeley Data (https://doi.org/10.17632/nrtmmbx3jg.1).
	The raw observation data will be provided upon requests to the corresponding author.
	The background images in Figure~\ref{fig:map} are provided by the Geospatial Information Authority of Japan.
	The XRAIN data were provided by by the Ministry of Land, Infrastructure, Transport and Tourism via the Data Integration and Analysis System (DIAS).

\end{acknowledgments}



\appendix

\section{Count-rate histories and individual energy spectra}\label{appendix}
\begin{figure*}[p]
	\begin{center}
	\includegraphics[width=0.8\hsize]{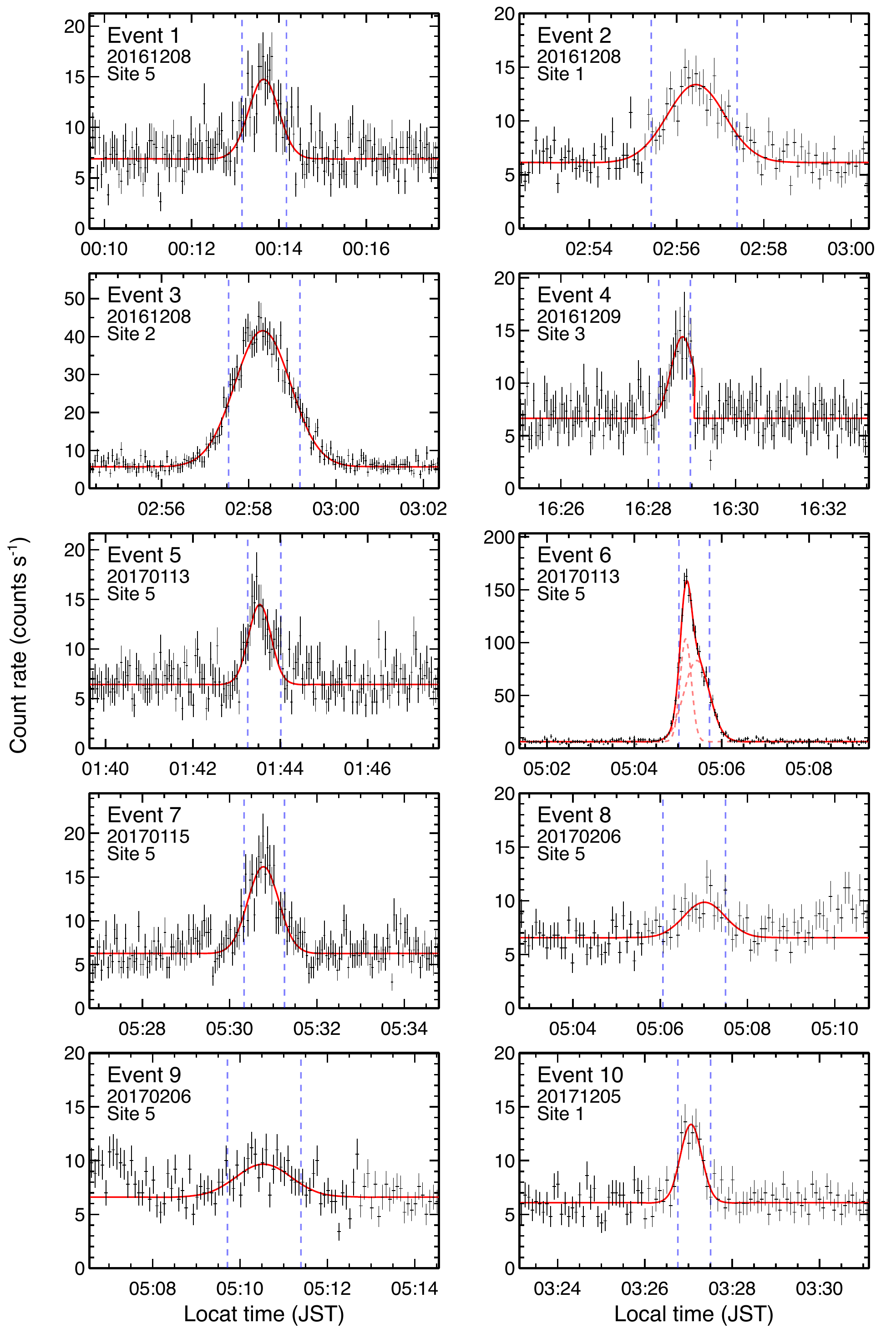}
	\caption{Figure~\ref{fig:lightcurve1}: Count-rate histories of Events~1--10 above 3~MeV. The local time (JST) is UTC+9~hours.
	The blue dashed lines indicate the start and end of the T80 duration. The red solid lines show the fitting results with the Gaussian function.
	If the histories are fitted by multiple Gaussian funcitons, the components are overlaid by red-dotted lines.
	The bin width is 3, 5, 3, 3, 3, 3, 3, 5, 5, and 5~sec for Events~1--10, respectively.}
	\label{fig:lightcurve1}
	\end{center}
\end{figure*}

\begin{figure*}[tbh]
	\begin{center}
	\includegraphics[width=0.8\hsize]{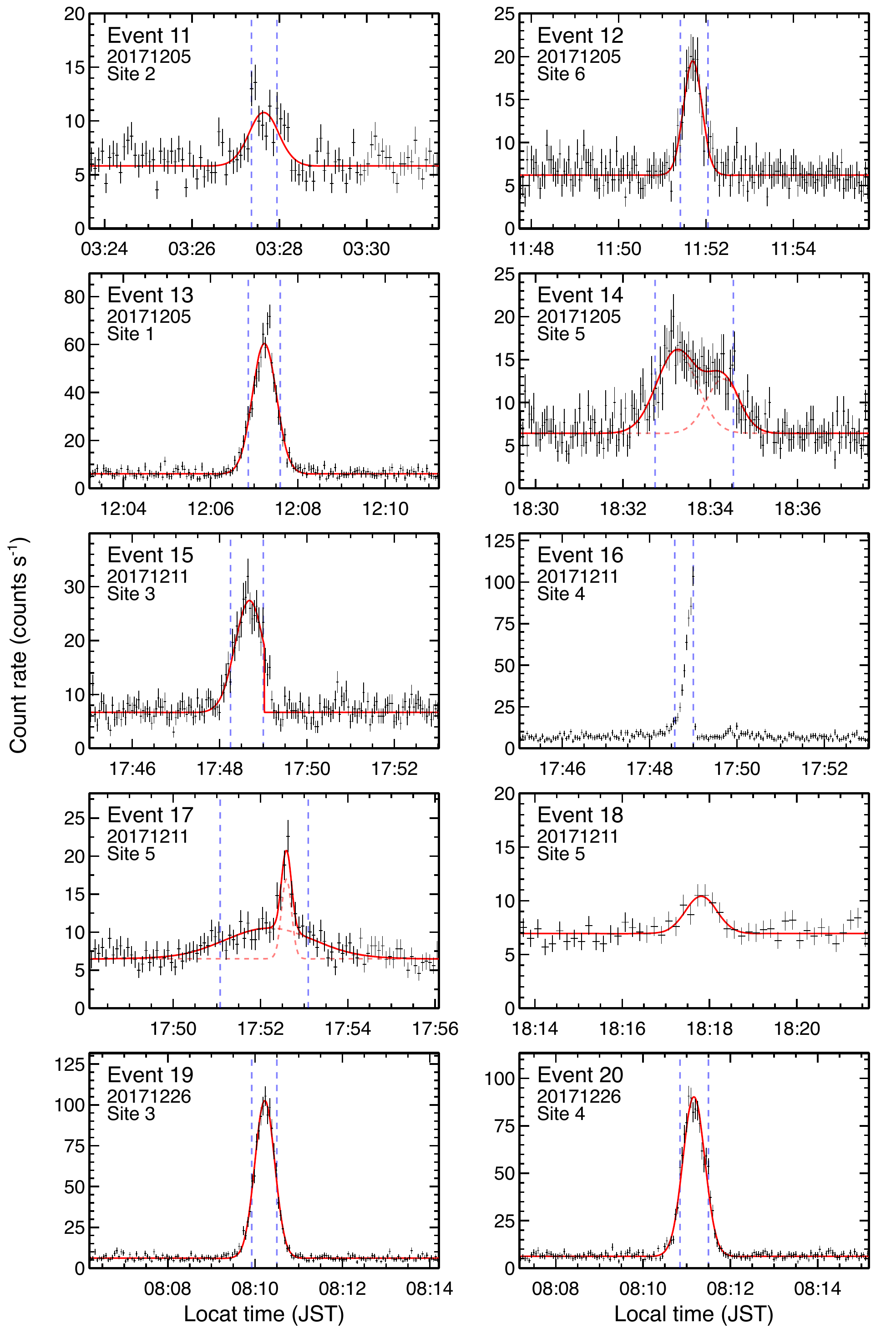}
	\caption{Figure~\ref{fig:lightcurve2}: Count-rate histories of Events~11--20 above 3~MeV in the same format as Figure~\ref{fig:lightcurve1}.
	The bin width is 5, 3, 3, 3, 3, 3, 5, 10, 3, and 3~sec for Events~11--20, respectively.}
	\label{fig:lightcurve2}
	\end{center}
\end{figure*}

\begin{figure*}[tbh]
	\begin{center}
	\includegraphics[width=0.8\hsize]{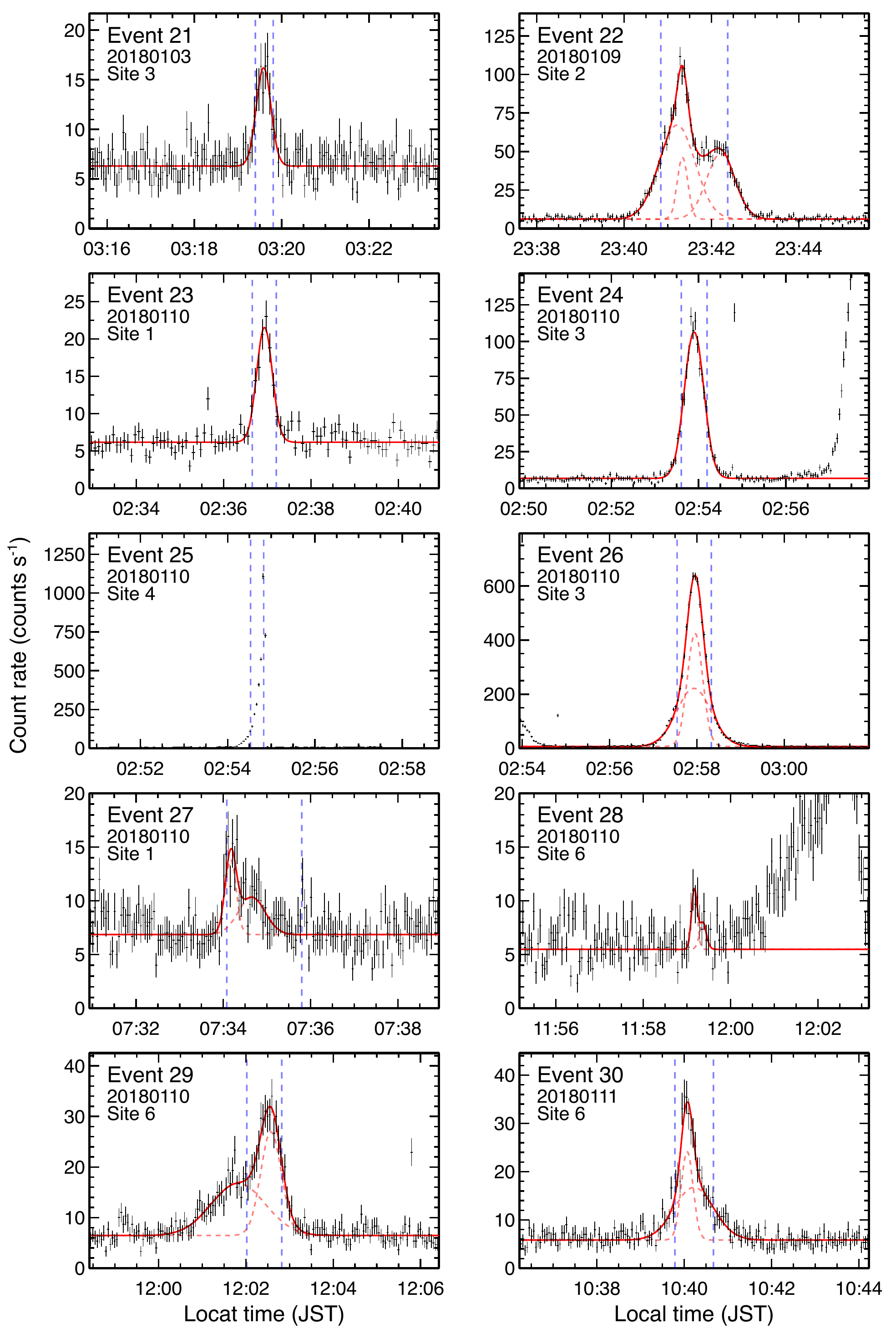}
	\caption{Figure~\ref{fig:lightcurve3}: Count-rate histories of Events~21--30 above 3~MeV in the same format as Figure~\ref{fig:lightcurve1}.
	The bin width is 3, 3, 5, 3, 3, 3, 3, 3, 3, and 3~sec for Events~21--30, respectively.}
	\label{fig:lightcurve3}
	\end{center}
\end{figure*}

\begin{figure*}[tbh]
	\begin{center}
	\includegraphics[width=0.8\hsize]{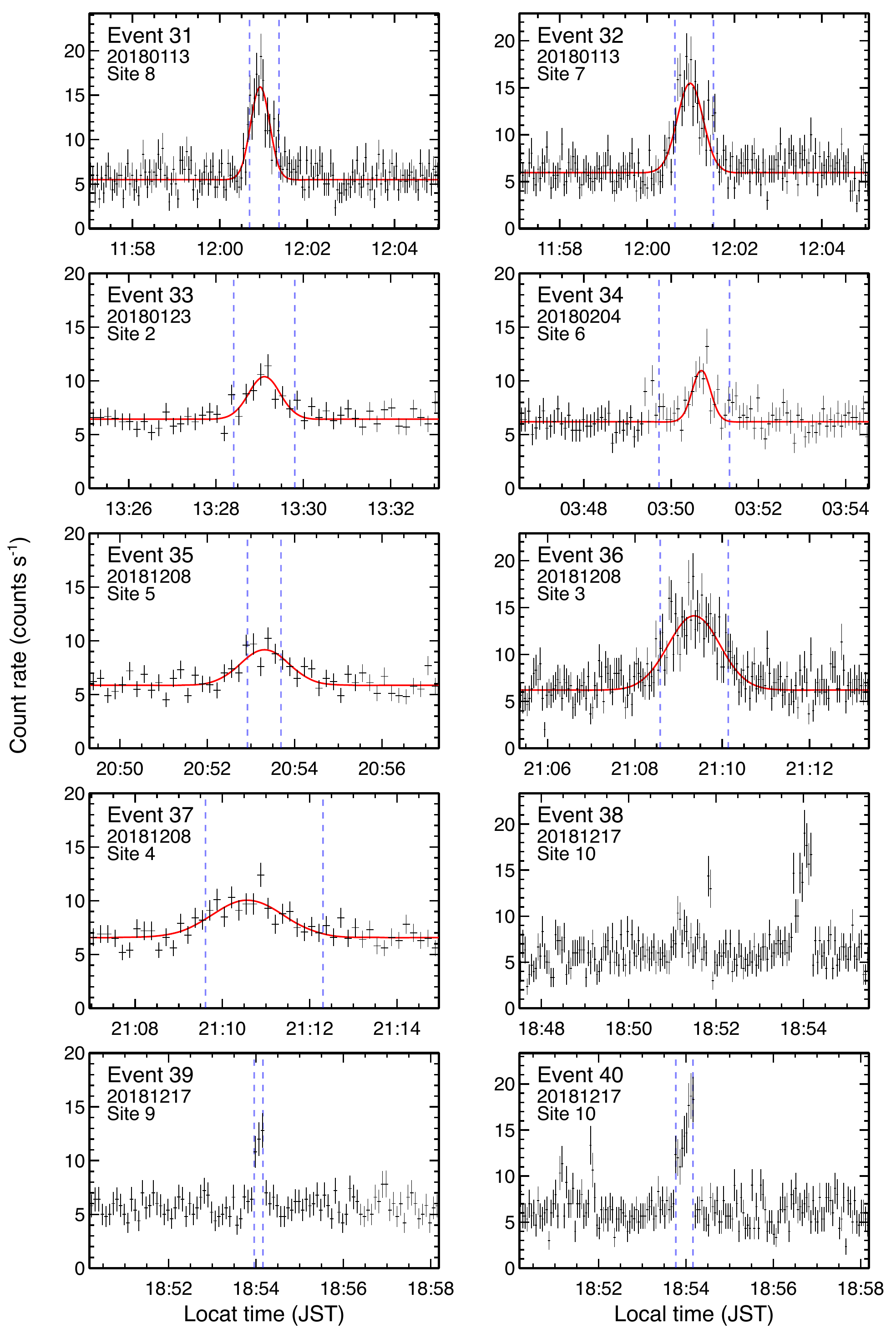}
	\caption{Figure~\ref{fig:lightcurve4}: Count-rate histories of Events~31--40 above 3~MeV in the same format as Figure~\ref{fig:lightcurve1}.
	The bin width is 3, 3, 10, 5, 10, 3, 10, 3, 5, and 3~sec for Events~31--40, respectively.}
	\label{fig:lightcurve4}
	\end{center}
\end{figure*}

\begin{figure*}[tbh]
	\begin{center}
	\includegraphics[width=0.8\hsize]{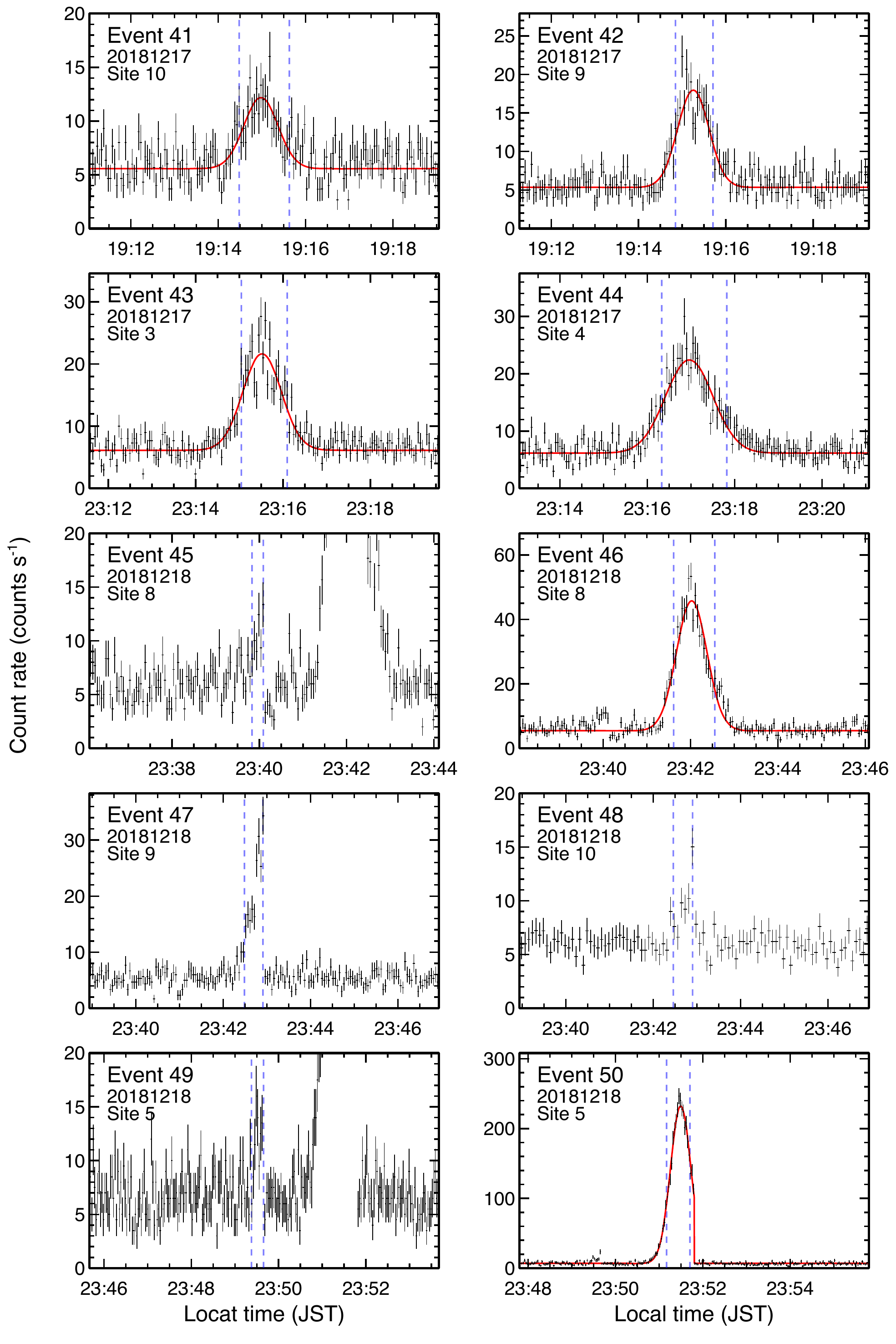}
	\caption{Figure~\ref{fig:lightcurve5}: Count-rate histories of Events~41--50 above 3~MeV in the same format as Figure~\ref{fig:lightcurve1}.
	The bin width is 3, 3, 3, 3, 3, 3, 3, 5, 2, and 2~sec for Events~41--50, respectively.}
	\label{fig:lightcurve5}
	\end{center}
\end{figure*}

\begin{figure*}[tbh]
	\begin{center}
	\includegraphics[width=0.8\hsize]{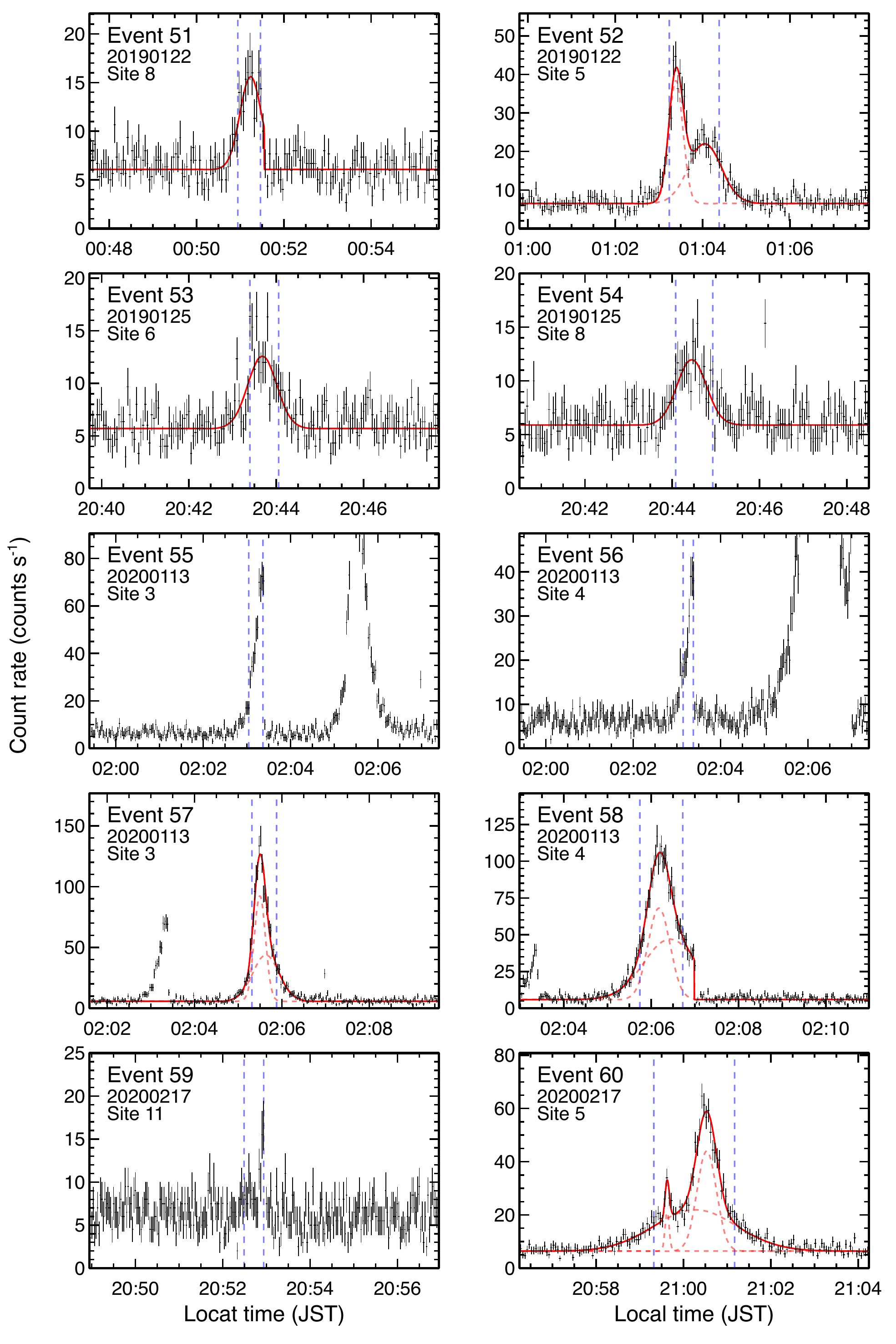}
	\caption{Figure~\ref{fig:lightcurve6}: Count-rate histories of Events~51--60 above 3~MeV in the same format as Figure~\ref{fig:lightcurve1}.
	The bin width is 3, 3, 3, 3, 2, 2, 2, 2, 2, and 3~sec for Events~51--60, respectively.}
	\label{fig:lightcurve6}
	\end{center}
\end{figure*}

\begin{figure*}[tbh]
	\begin{center}
	\includegraphics[width=0.8\hsize]{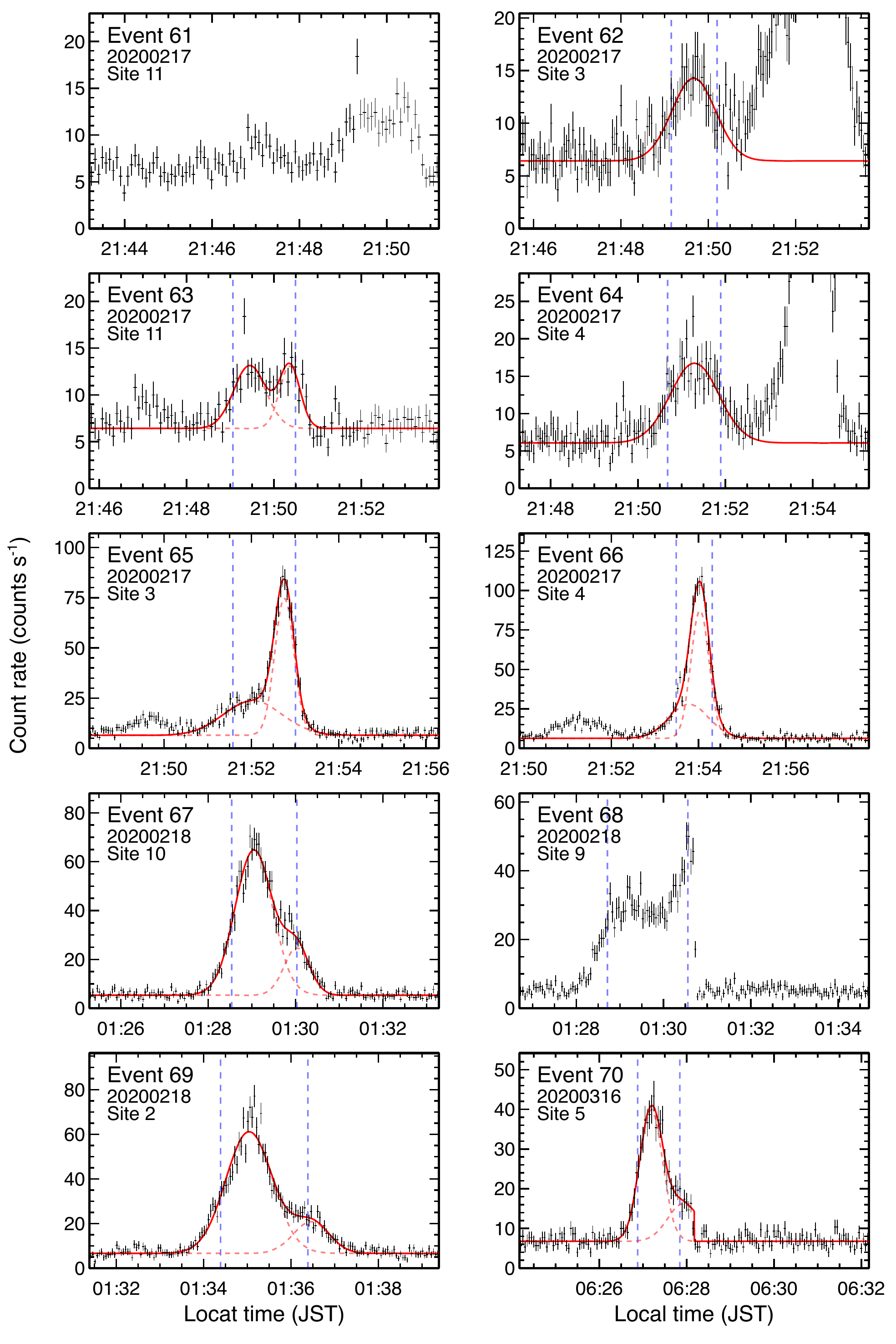}
	\caption{Figure~\ref{fig:lightcurve7}: Count-rate histories of Events~61--70 above 3~MeV in the same format as Figure~\ref{fig:lightcurve1}.
	The bin width is 5, 3, 5, 3, 3, 3, 3, 3, 3, and 3~sec for Events~61--70, respectively.}
	\label{fig:lightcurve7}
	\end{center}
\end{figure*}

\begin{figure*}[p]
	\begin{center}
	\includegraphics[width=0.8\hsize]{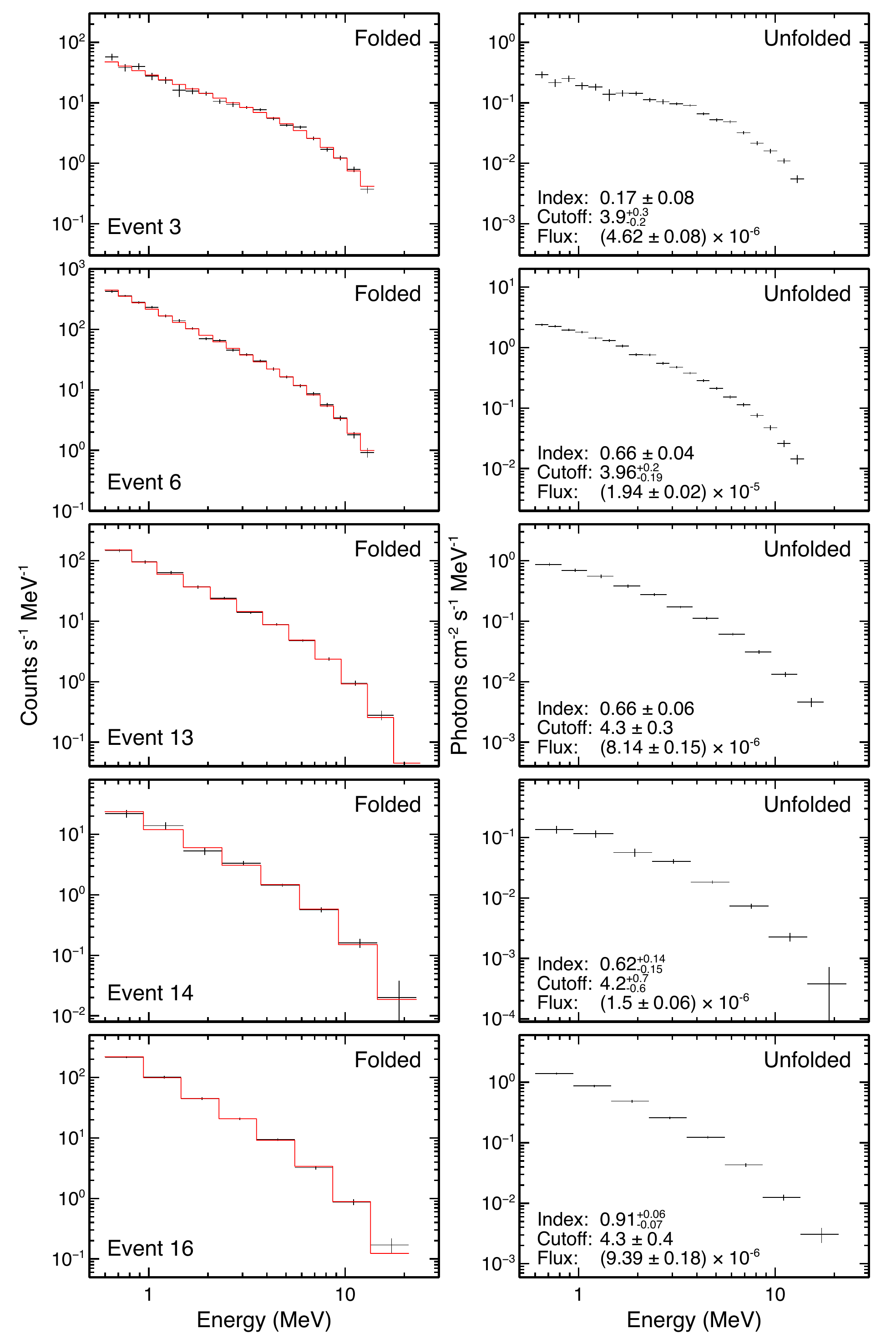}
	\caption{Figure~\ref{fig:spectrum1}: Energy spectra of Events 3, 6, 13, 14, and 16. 
	The left panels are in the count-rate space, and the right panels are in the photon-flux space produced by unfolding detector responses with {\tt XSPEC}.
	The best-fit models are overlaid by the red-solid lines. The units of the cutoff energy and energy flux are MeV and erg~cm$^{-2}$~s$^{-1}$, respectively.}
	\label{fig:spectrum1}
	\end{center}
\end{figure*}

\begin{figure*}[tbh]
	\begin{center}
	\includegraphics[width=0.8\hsize]{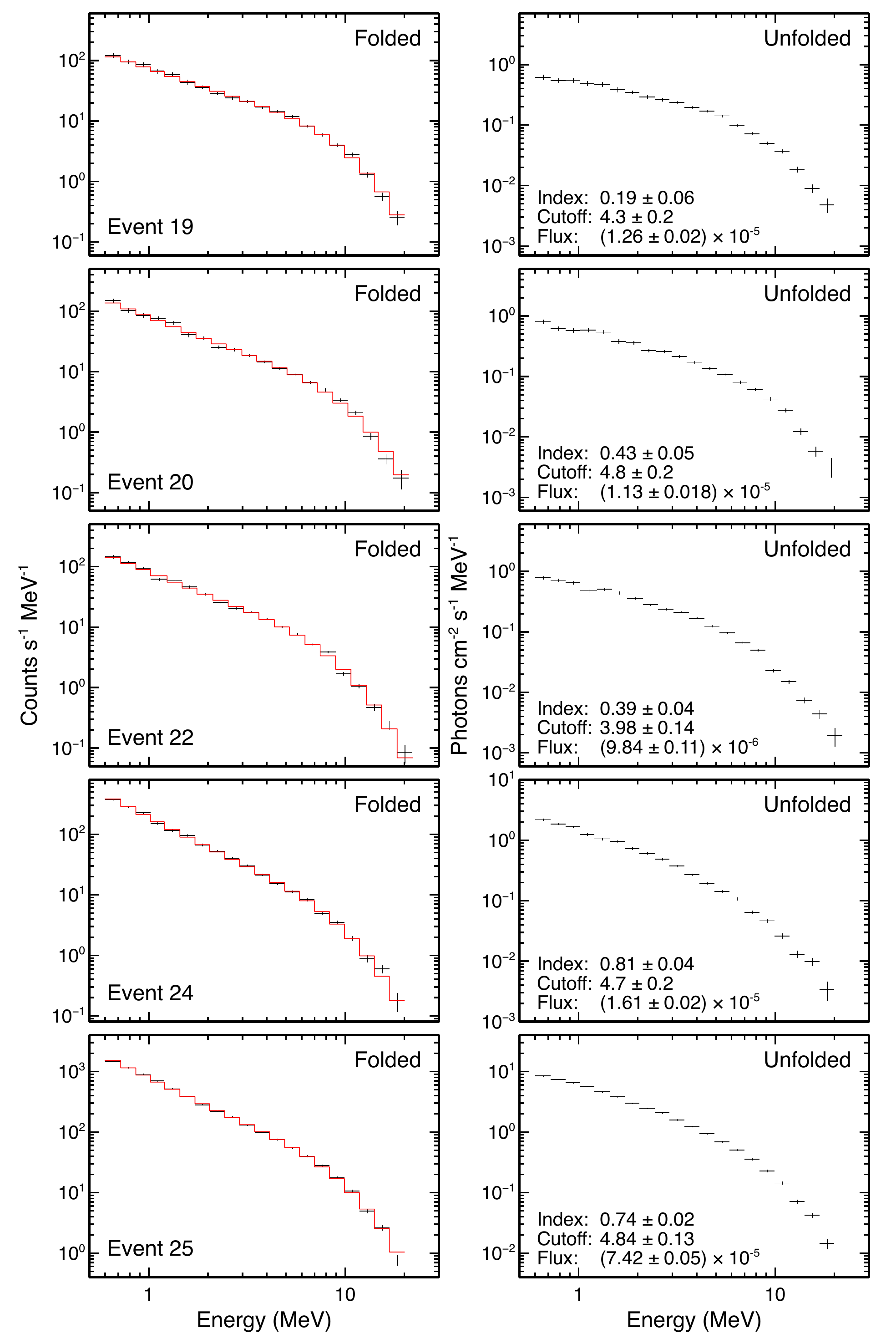}
	\caption{Figure~\ref{fig:spectrum2}: Energy spectra of Events 19, 29, 22, 24, and 25 in the same format as Figure~\ref{fig:spectrum1}.}
	\label{fig:spectrum2}
	\end{center}
\end{figure*}

\begin{figure*}[tbh]
	\begin{center}
	\includegraphics[width=0.8\hsize]{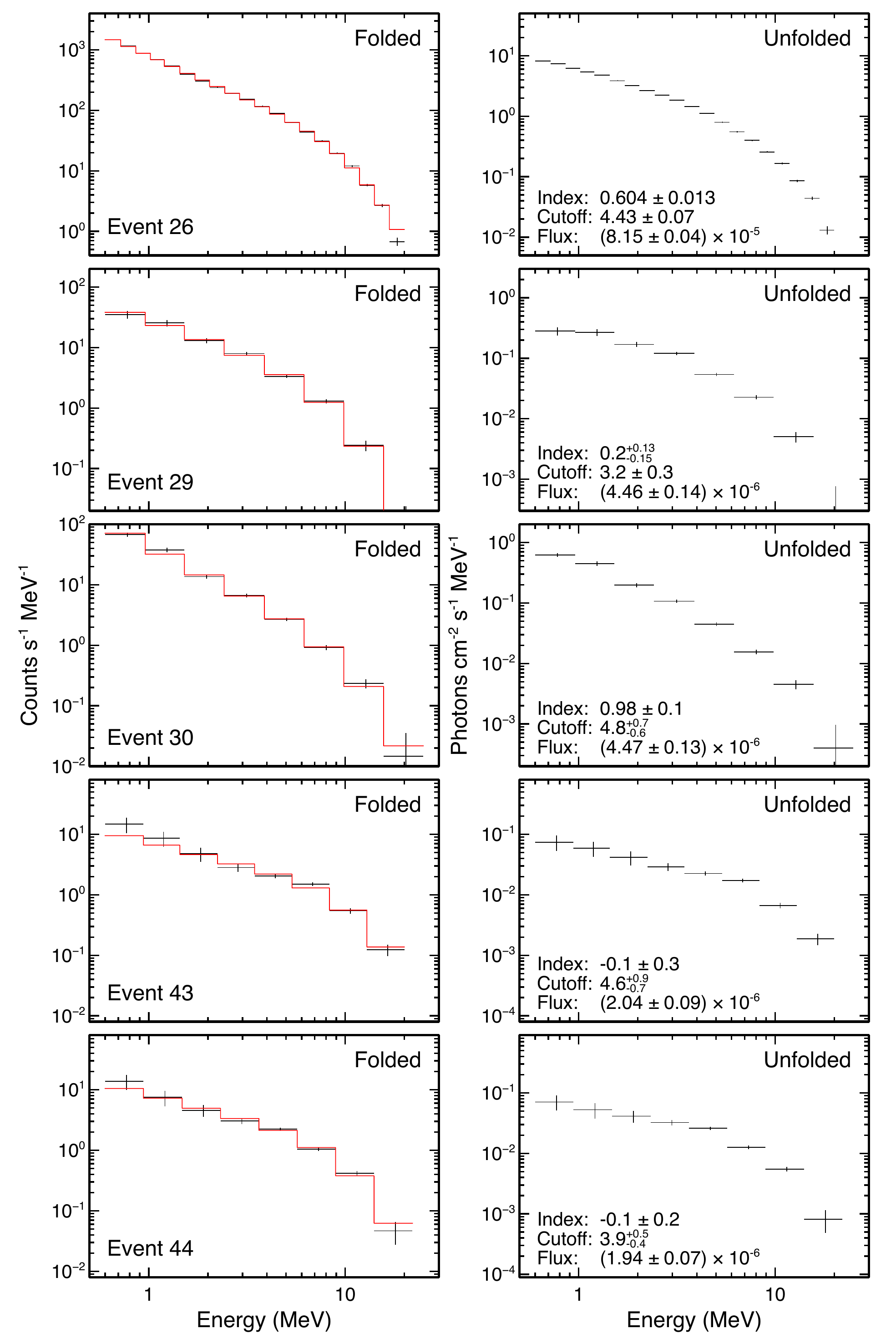}
	\caption{Figure~\ref{fig:spectrum3}: Energy spectra of Events 26, 29, 30, 43, and 44 in the same format as Figure~\ref{fig:spectrum1}.}
	\label{fig:spectrum3}
	\end{center}
\end{figure*}

\begin{figure*}[tbh]
	\begin{center}
	\includegraphics[width=0.8\hsize]{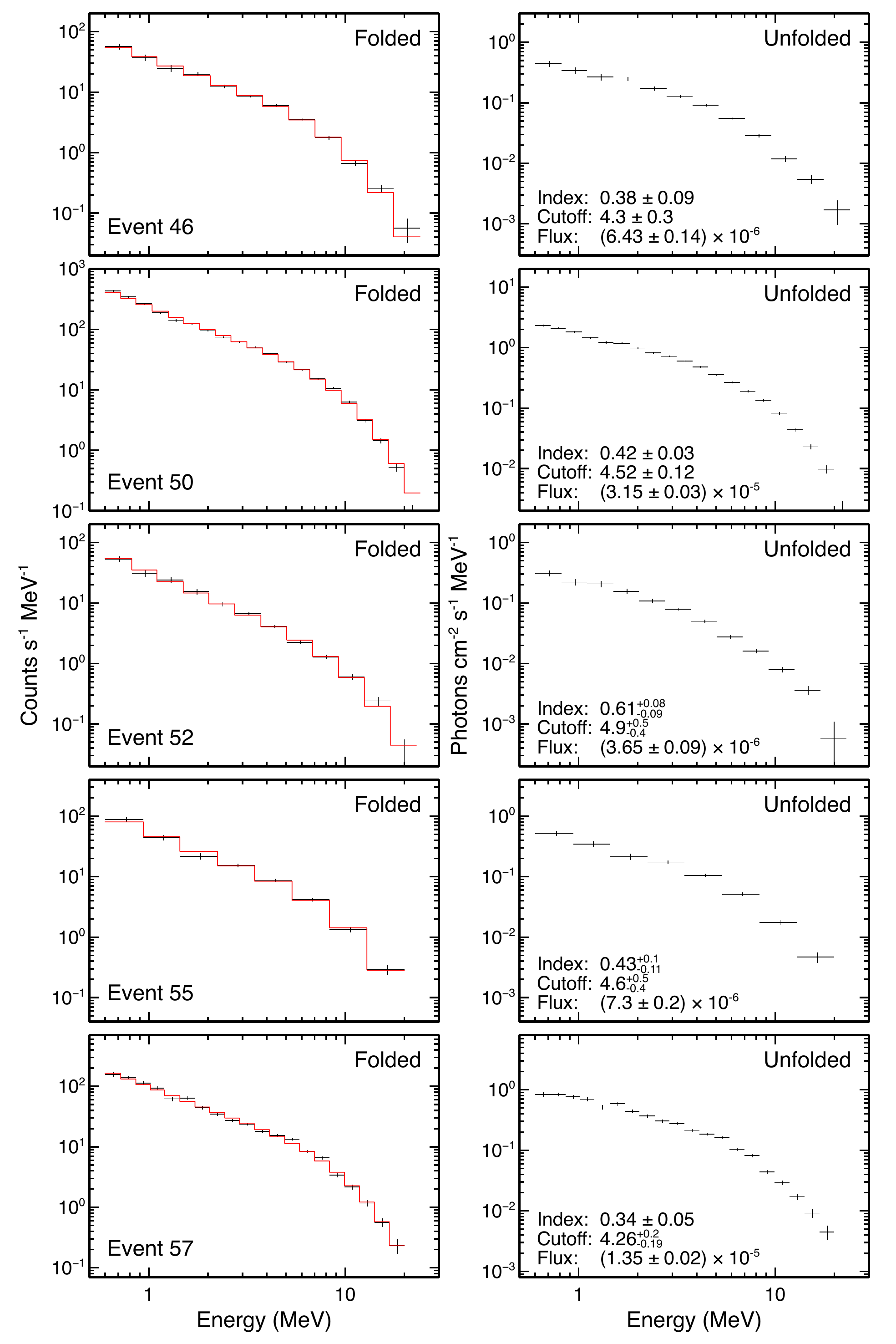}
	\caption{Figure~\ref{fig:spectrum4}: Energy spectra of Events 46, 50, 52, 55, and 57 in the same format as Figure~\ref{fig:spectrum1}.}
	\label{fig:spectrum4}
	\end{center}
\end{figure*}

\begin{figure*}[tbh]
	\begin{center}
	\includegraphics[width=0.8\hsize]{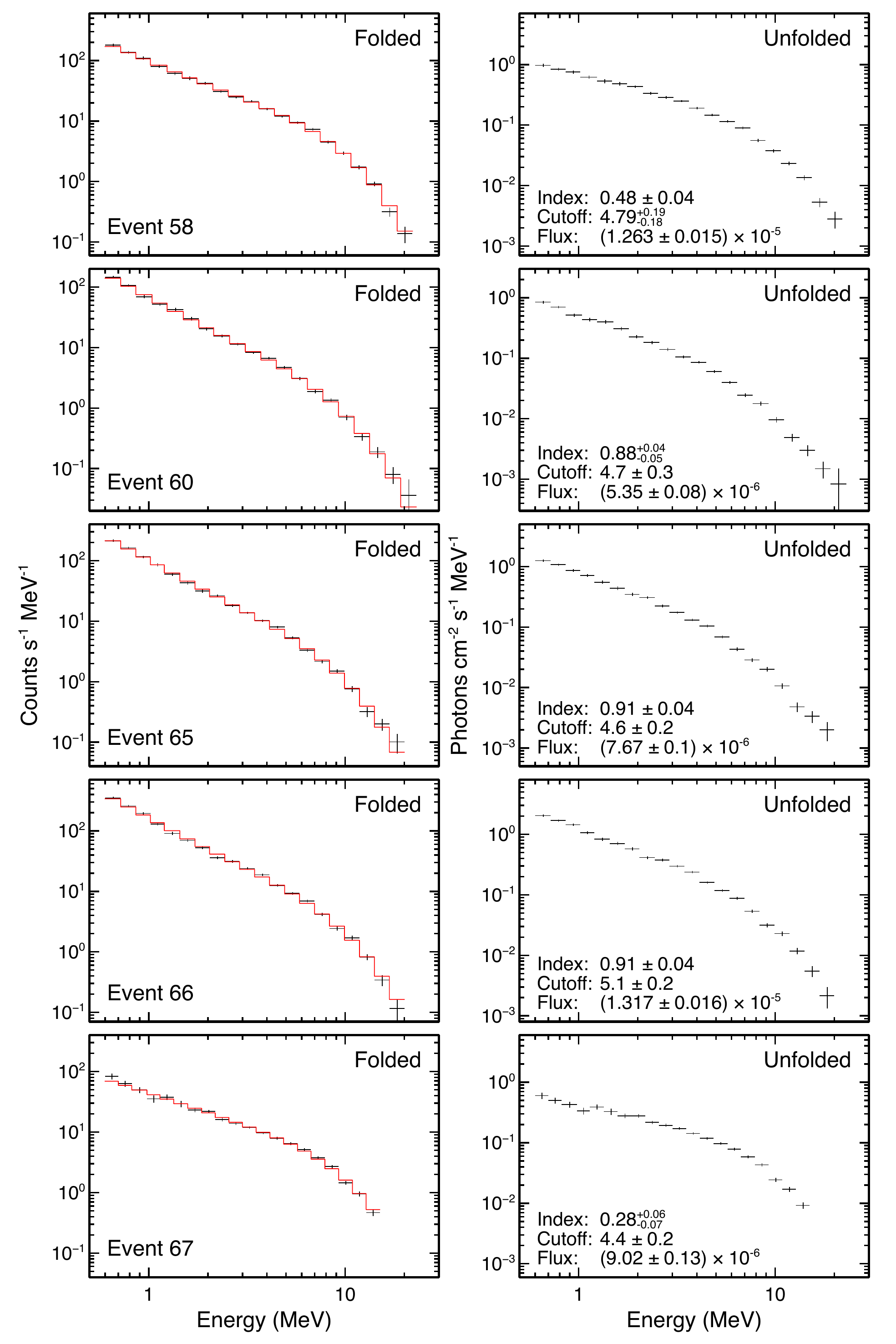}
	\caption{Figure~\ref{fig:spectrum5}: Energy spectra of Events 58, 60, 65, 66, and 67 in the same format as Figure~\ref{fig:spectrum1}.}
	\label{fig:spectrum5}
	\end{center}
\end{figure*}

\begin{figure*}[tbh]
	\begin{center}
	\includegraphics[width=0.8\hsize]{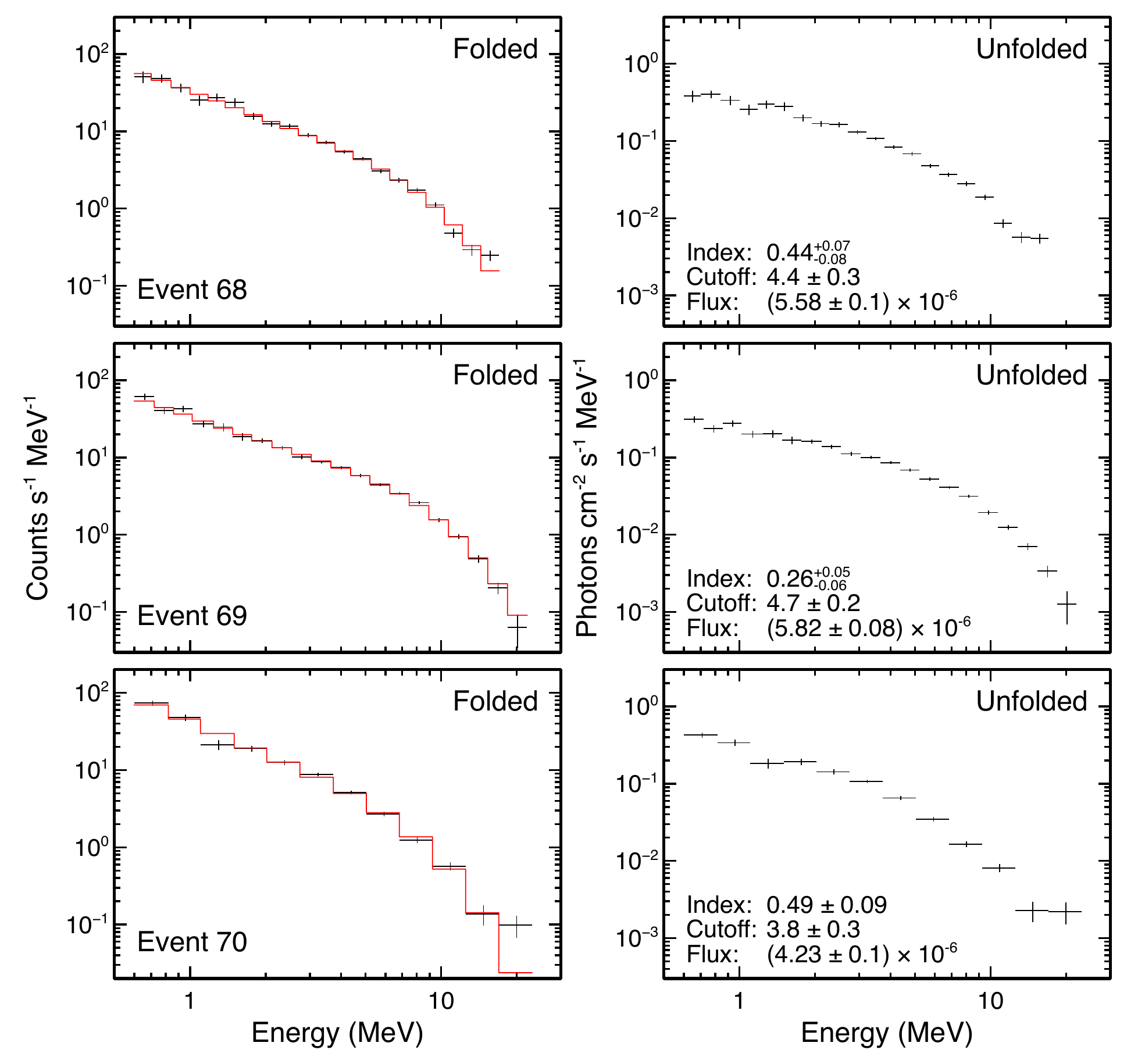}
	\caption{Figure~\ref{fig:spectrum6}: Energy spectra of Events 68, 69, and 70 in the same format as Figure~\ref{fig:spectrum1}.}
	\label{fig:spectrum6}
	\end{center}
\end{figure*}

	As a gamma-ray glow catalog, count-rate histories of all the detected glows above 3~MeV are listed in Figure~\ref{fig:lightcurve1} to Figure~\ref{fig:lightcurve7}.
	Also energy spectra of 28 bright glows are shown in Figure~\ref{fig:spectrum1} to Figure~\ref{fig:spectrum6}, with response-folded and response-unfolded forms.
	All the data points are disclosed on the public data archive for the scientific community and the general public as mentioned in the Acknowledgments.


%

\clearpage

\end{document}